\documentstyle[11pt,supertab]{article}
\def\be{\begin{equation}}

\def\ee{\end{equation}}

\def\eg{{\em e.g.}\ }

\def\dg{\mbox{$^\circ$}}		%degree sign
\def\etal{{\em et al.}}

\def\hMpc{h^{-1}{\rm Mpc}}
\def\h3Mpc{h^{-3}{\rm Mpc}^3}

\def\h3Mpcinv{h^{3}{\rm Mpc}^{-3}}
\def\ie{{\em i.e.}\ }

\def\rms{\mbox{\em rms}}
\def\ten#1{\times 10^{#1}}
\def\veps{\varepsilon}

\def\wth{\mbox{$w(\theta)$}}

\def\spose#1{\hbox to 0pt{#1\hss}}
\def\simlt{\mathrel{\spose{\lower 3pt\hbox{$\mathchar"218$}}
     \raise 2.0pt\hbox{$\mathchar"13C$}}}
\def\simgt{\mathrel{\spose{\lower 3pt\hbox{$\mathchar"218$}}
     \raise 2.0pt\hbox{$\mathchar"13E$}}}

\def\inserthang{\everypar={\parindent=0pt\hangindent=1.5pc\hangafter=1}}
\def\refmode{\parindent=0pt\inserthang\parskip 0pt
\frenchspacing\exhyphenpenalty=10000\hyphenpenalty=10000}
\def\refer#1{#1}
\def\refeq#1{\relax}		% for 'quiet' ref
\def\IAU130{in IAU Symp. 130, Large Scale Structures of the Universe}

\input epsf

\def \PRElim{$2.05 \times 10^{-2}$}

\begin{document}

\begin{center}
 {\Large\bf The APM Bright Galaxy Catalogue}\\[0.5in]

 {\bf Jon Loveday\\[0.1in]
      Fermi National Accelerator Laboratory,\\
      PO Box 500, Batavia,\\
      Illinois 60510, USA.\\[0.2cm]
      loveday@fnal.gov}\\[0.5cm]
	Revised August 23, 1995
\end{center}

\section*{Abstract}

The APM Bright Galaxy Catalogue lists positions, magnitudes, shapes and
morphological types for 14,681 galaxies brighter than $b_J$ magnitude 16.44
over a 4,180 square degree area of the southern sky.
Galaxy and stellar images have been located from glass copy plates of the 
United Kingdom Schmidt Telescope (UKST)
IIIaJ sky survey using the Automated Photographic Measuring (APM) facility in
Cambridge, England.
The majority of stellar images are rejected by the regularity of their 
image surface brightness profiles.
Remaining images are inspected by eye on film copies of the
survey material and classed as stellar, multiple stellar, galaxy, merger or
noise.
Galaxies are further classified as elliptical, lenticular, spiral, irregular 
or uncertain.
The 180 survey fields are put onto a uniform photometric system by comparing
the magnitudes of galaxies in the overlap regions between neighbouring plates.
The magnitude zero-point, photometric uniformity and photographic saturation
are checked with CCD photometry.
Finally, the completeness and reliability of the catalogue is assessed
using various internal tests and by comparing with several independently
constructed galaxy catalogues.

{\bf Key words:} catalogues --- galaxies: fundamental parameters ---
galaxies: general --- galaxies: photometry.

\section{Introduction}

The APM Galaxy Survey (Maddox \etal\ 1990a, b) includes about
two million galaxies down to magnitude $b_J = 20.5$ over 4300
square degrees of the southern sky.
It was the first machine-generated galaxy survey to cover an area
of sky significantly larger than one Schmidt plate, and has proved
an important survey for measuring galaxy number-magnitude counts
over a wide magnitude range (\refer{Maddox \etal\ 1990c}) and particularly
for the most reliable measurement to-date of the angular correlation
function of galaxies on large scales (\refer{Maddox \etal\ 1990d}).
This latter measurement was one of the first results to rule out the
standard cold dark matter model of galaxy formation 
(eg. \refer{Davis \etal\ 1985}).

Unfortunately, the APM Galaxy Survey, while complete to a faint magnitude
limit of $b_J = 20.5$, is not very reliable for galaxies brighter than
$b_J \approx 16.5$.
There are several reasons for this.
Firstly, the surface density of galaxies brighter than $b_J \approx 16.5$ 
is only
about $1/20$ of the surface density of stars at the same magnitude limit
even at the galactic poles.
Therefore the selection of an uncontaminated bright galaxy sample requires
an exceptionally reliable method of rejecting stars and merged images.
Secondly, photographic emulsions have a limited dynamic range,
and in order to detect images as faint as $b_J = 20.5$, the brighter
images are necessarily saturated.
Thirdly, bright stars have diffraction spikes and `ghost' images 
(\refer{UKSTU handbook})
and large galaxies contain sub-structure.
All of these factors prevent the standard APM image parameters,
which were designed to classify small, faint images,
from selecting a sufficiently reliable bright galaxy catalogue.

This paper describes the construction of a bright galaxy catalogue,
complete to $b_J = 16.44$, using the same APM scans used for the faint survey.
We developed a semi-automated method of star-galaxy
separation, whereby most stellar images were rejected (losing only
about 3\% of galaxies) and the remaining images inspected by eye
on a film copy of the photographic plate.
The distinction is emphasized between a 
survey {\em constructed} by eye, for example 
the \refer{Zwicky \etal\ (1961-68)}
catalogue or the Lick (\refer{Shane and Wirtanen 1967}) survey,
where the observer has
to locate each image and {\em then} decide whether it should be included
in the catalogue, and a semi-automated survey like the present one
where the observer is
given the position of each image satisfying a magnitude limit,
and then classifies it as a galaxy or star.
It is much easier for the eye 
to distinguish a galaxy from a star than it is to select a complete magnitude
or diameter limited sample,
and so the semi-automated survey should be much more reliable.

The APM Bright Galaxy Catalogue (APM-BGC) covers almost the same area
as the faint APM galaxy survey of Maddox \etal\ (1990a, b), including
180 out of the 185 fields of the fainter survey, an area of approximately
4,180 square degrees.
Figure~\ref{fig:fields} shows the 
distribution of the 180 survey fields in an equal area projection on the sky.

The construction of the present catalogue was first described by 
Loveday (1989).
A similar survey has been carried out by Raychaudhury (1989)
in the region towards the `Great Attractor', and there is an ongoing effort
(Raychaudhury \etal\ 1994) to map out the galaxy distribution near
the equator.

The layout of the paper is as follows.
The construction of the bright galaxy catalogue, including star-galaxy
separation and plate matching, is described in \S\ref{sec:constr}.
Internal tests for uniformity, completeness and consistency are described in
\S\ref{sec:int_tests} and comparison with other catalogues is made
in \S\ref{sec:other_cats}.
\S\ref{sec:photometry} describes the CCD calibrations used to check
the APM to $b_J$ magnitude conversion, as a further
test of photometric uniformity in the survey, and to define a second-order
correction for photographic saturation.
In \S\ref{sec:catalogue} we describe the catalogue data and
present plots of the galaxy distribution.
We compare the angular and spatial correlation functions
of early and late type galaxies in the APM-BGC in \S\ref{sec:clust}.
Finally, the properties of the catalogue are summarized in 
\S\ref{sec:summary}.

\section{Construction of the Catalogue}
\label{sec:constr}

\subsection{The APM Measurements}

The Automated Plate Measuring (APM) machine in Cambridge
is a high-speed laser microdensitometer with on-line
image detection and processing.
Technical details are given by 
\refer{Kibblewhite \etal\ (1984)} and by \refer{Maddox (1988)}.
Plate scanning occurs in two passes---firstly the sky background is
measured in $640 \times 640$ 0.5mm (33.6 arcsec) square pixels over the plate,
with images being removed by median filtering.
Note that if a galaxy image occupies a substantial fraction of a sky pixel,
then the sky will be biased high in that pixel and hence the flux from
that and nearby galaxies will be underestimated.
This could cause us to bias against low surface brightness galaxies.
We plan to investigate this possible bias using simulations in a future paper.

In the second pass, images are detected as
connected groups of pixels with densities higher than
a set threshold above local sky, determined from bilinear interpolation
of the background map.
Each image is parameterized by fifteen numbers: integrated isophotal density 
$D$; $x$ and $y$ coordinates on the plate measured in $8\mu$ units;
three second-order moments $\sigma_{xx}$, $\sigma_{xy}$ and $\sigma_{yy}$;
peak density; and finally an areal profile---the image area $A_i$ above eight
density levels $D_i = t + 2^{(i-1)}$,
where $i$ runs from 1 to 8.
The threshold $t$ was set at twice the \rms\ noise in the measured sky,
and corresponds to a surface brightness of 
$\mu_J \approx 24.5$ -- $25 b_J$ mag arcsec$^{-2}$ 
\refer{(Maddox \etal\ 1990a)}.
The APM magnitude $m$ is defined as $+2.5 \lg D$, and
so is reversed in sign from usual magnitude systems.
A rough conversion, accurate to $\approx$ 0.5 mag, from $m$ to $b_J$ 
for $b_J \simlt 17$ is given by $b_J \approx 29 - m$.
This zero-point differs from that for faint images
due to mild saturation of the brighter galaxies.

\subsection{Star-Galaxy Separation}

For twelve of the survey plates, a $90 \times 90$ pixel raster scan
was made, using the APM machine, around the 4000 brightest images on each 
plate.
Displaying each raster scan in turn enabled rapid classification
of these images.  
Unfortunately, constraints on the amount of APM scanning time
available meant that only a small fraction
of the survey plates could be scanned in this way.
Therefore, a semi-automated method for separating stars from galaxies,
using just the standard APM image parameters described above,
was devised and is described in this subsection.

In the faint survey (Maddox \etal\ 1990a), areal profile
information (the `$\psi$-classifier') is used
to discriminate galaxy from stellar and noise images on each survey plate.
The quantity $\psi$ is defined as
\be
  \psi = 1000 \lg \left( \sum_{i=1}^{10} w_i(m)[p_i(m) - p_i^*(m)]^2 \right),
  \label{eqn:psi}
\ee
where $p_i$ is the $i$th areal profile for $i=1 \ldots 8$, the peak density
for $i=9$ and the radius of gyration for $i=10$.
For each parameter $p_i$, a scatter plot against magnitude $m$ is produced,
and the stellar locus $p_i^*(m)$ is located as a function of magnitude.
For each image the difference between the parameter $p_i$ and the
stellar locus $p_i^*(m)$ at the appropriate magnitude is calculated
for all $i$.
The differences from each $p_i$ are summed in quadrature with a set
of weighting factors $w_i$, equal to the reciprocal of the estimated
variance in $p_i$.

For objects in the magnitude range $17 < b_J < 20.5$, selecting 
images with $\psi > 1000$
yields a galaxy
sample that is $\approx 95\%$ complete and with $< 10\%$ contamination
from stellar and merged stellar images.
For brighter objects the number of stars relative to the number of
galaxies becomes very large and so a $\psi$-selected galaxy sample will
suffer from an unacceptable rate of contamination.

At $b_J \approx 16$, the best galaxy sample that can be selected using
the $\psi$ parameter has $>50\%$ contamination from stars and merged images,
although many of the merged objects can be identified using an additional two
parameters based on the radius of gyration and the fraction of
saturated image area (Maddox \etal\ 1990a).
Use of the `$k$' and `$\mu$' parameters reduces merger contamination
to $\sim 5\%$ at $b_J \approx 17$.

For such bright objects, the APM
parameterization no longer contains all the information available from the
UKST plates.
The stars display haloes and diffraction spikes,
and galaxies have detailed sub-structure.
Also, both stars and galaxies are saturated.
Therefore we used areal profile information in
conjunction with visual inspection of photographic plate copies
in order to select a more reliable bright galaxy sample.

The 3000 or so images with $m > 12.0$ ($b_J \simlt 17.0$)
on each plate from the scans of Maddox \etal\ (1990a,b)
are sorted by density and binned with twenty objects per bin.
Stellar profiles
cluster closely around the median for each bin as $\approx 95\%$ of images with
$b_J \le 16.5$ are stars.  Since galaxies are resolved,
extended objects, they have a broader profile and thus stand out from
the stars.
Figure~\ref{fig:profile} shows areal profiles for (a) 10\% of stars and (b)
all galaxies in the APM magnitude range 12.5--12.6 on one plate.
The median area is
calculated at each profile level in each magnitude bin.  
Twenty objects per bin is a good compromise between having
a small density range per bin but
therefore only a few images to determine the median, and having wider
bins giving a less noisy median but wider scatter about the median.

We define the `profile residual error' (PRE), $\varepsilon_j$, for image $j$ by
\begin{equation}
  \label{eqn:pre}
  \varepsilon_j = \langle A^1_{m(j)}\rangle^z \sum_{i=1}^8 {{{|A_j^i - 
	  	  \langle A^i_{m(j)}\rangle|}^x} \over 
          	  \langle A^i_{m(j)}\rangle^y}.
\end{equation}
Here $A_j^i$ is the area at the $i$th profile level for image $j$ and 
$\langle A^i_{m(j)}\rangle$ the
median area at the $i$th profile level for $j$'s magnitude bin.
The denominator
$\langle A^i_{m(j)}\rangle^y$ gives extra weight to higher profile levels,
where there is smaller area.
This is desirable since at low surface brightness levels haloes around bright
stars often resemble galaxies, it is only towards the brighter core of the 
stellar image that bright stellar profiles become regular.  
The factor $\langle A^1_{m(j)}\rangle^z$ scales each PRE by the median 
isophotal area for that magnitude bin to the power $z$,
since one expects large images, having more pixels, to suffer less
from Poisson noise.   This definition of profile residual error is more general
than the $\psi$ classifier (equation~\ref{eqn:psi})
used for the faint survey (\refer{Maddox 1988},
Maddox \etal\ 1990a)
and allows a better star/galaxy separation for bright images to be achieved.
Choice of the parameters $x$, $y$, $z$ is discussed below.

The raster scan classifications proved to be very valuable for calibrating
the performance of the PRE in separating stars from galaxies.
Figure~\ref{fig:pre} plots the profile residual error $\veps$ against APM
magnitude for one field.
The dashed horizontal line marks the PRE cutoff used for rejecting stars
and the symbols show image classifications made from the raster scans.
It is clear from Figure~\ref{fig:pre}, that even using the customized PRE 
parameter, visual checking of some
images is unavoidable to obtain a $\simgt 95\%$ complete galaxy sample
with less than 5\% stellar contamination.

It was decided to aim for a galaxy sample with $97\%$ completeness and with
negligible stellar contamination by rejecting obvious stars with a
low value of $\veps$ and then inspecting the remaining images on a film
copy of the plate.
The weighting powers $x, y, z$ were adjusted to minimize the number of
images $N_{eye}$
that needed to be inspected to obtain a completeness of 97\% for the twelve
plates which had been pre-classified from
raster scans.
The optimal values of $x, y, z$ were found to be $x = 1.00$,
$y = 2.14$, $z = 0.675$.  
By minimizing $N_{eye}$ for each of the twelve rastered fields
separately, the parameters $x,y,z$ were found to have standard
deviations of 0.2, 0.4 and 0.2 respectively.
Using these parameters required all images with $\veps > $\PRElim,
an average of 539 objects for each of these twelve plates, 
to be inspected to obtain 97\%
completeness.
An aimed completeness of 98\% would have required an average of 665 objects
to be inspected on each plate.
As well as taking about 20\% longer, nearly all of the additional images
would be stars, and the `tedium factor' of checking so many stars might
actually cause more galaxies to be missed.

There is a danger that plate-to-plate variations in $\veps$ (\eg due
to differences in seeing) could
give rise to changes in completeness, but this should be unimportant since
we identify only about 125 galaxies per plate
down to $b_J \approx 16.5$, and so a change in completeness of 3\%
would have a much smaller effect on the number of identified galaxies
than Poisson fluctuations in the galaxy density.

\subsection{Plate Eyeballing}

\subsubsection{Masking Large Images} \label{sec:large_ims}

Images larger than 1--2mm in diameter cause
problems since they are often either removed by the APM machine
or split up into many sub-images.
In order to understand why the APM machine removes large images, 
some understanding
of the way in which APM scanning is carried out is necessary.
Note that since the plates for this survey were scanned, the APM hardware
has been upgraded which completely avoids the problem.
The following description refers to the operation of the APM machine when 
the APM-BGC survey was being carried out.

The APM machine scans the plate in 3 areas---horizontal strips across the top,
middle and bottom of the plate.
Each area is scanned in 2mm columns, each of which overlaps the previous
column by 1mm, so that nearly every part of the plate is scanned twice,
in addition to the preliminary background scan.
The pixel locations and densities above the
threshold in each column are stored in memory so that
they can be merged with adjoining pixels in the subsequent column.
Due to available memory constraints, image pixels cannot be stored for
more than one column, and so
any group of pixels spanning more than two scan columns is dropped
from the image list.
Thus all images larger than 2mm in horizontal extent are lost,
and those larger than 1mm will be lost if they happen to span three columns.
As an example, the dense central region of a globular cluster
will often be removed,
leaving an annulus of tightly-packed images marking the outer regions of the 
cluster.
Also, large images will of course obscure any galaxies lying behind them.
Thus regions around large images were `drilled' out of the survey as
follows.

A grey-scale plot of the APM background map was made for each field
and compared with
a film copy of the plate.  Any large images and bad satellite trails were
marked on the plot.  The same background map was then displayed on an image
display and the vertices of parallelogram-shaped holes defined interactively.
Parallelograms were chosen to enable satellite trails to be removed,
although in practice these trails rarely showed up on the background maps
and most holes drilled were rectangular.  A file of hole coordinates was
saved, with typically ten holes drilled per plate.

\subsubsection{Image Classification}

All objects brighter than $m = 12.0$, an average of 3124 images per plate,
were extracted from $\sim 300,000$ images detected by each scan.
The profile residual error $\veps$ (equation~\ref{eqn:pre}) 
was calculated for each of these bright images.
Areas around large images were drilled out as described above,
and a plate-scale finding chart 
marking objects brighter than $m = 12.5$ and with PRE $\veps >$ \PRElim,
425 objects per plate when averaged over the whole survey, was made.
By placing the finding chart under a film copy of the plate,
these objects were inspected with a
magnifying lens and assigned a 1 digit classification code as
shown in Table~\ref{tab:classns}.
Blended images were only placed into class 8 when stellar and
galactic components were of
comparable intensity as judged by eye.  Most large galaxies contain one or
two faint stellar images which were ignored in this classification scheme.
An average of 124 objects per plate (\ie about 30\% of those inspected) 
were classified as a galaxy, and eyeballing each plate took about two hours.

\subsection{Plate Matching} \label{sec:matching}

It is important for statistical studies of galaxy clustering that a
catalogue be uniform, or at least have a known selection function, over
its whole area.
Due to slight variations in observing, processing and scanning conditions,
the magnitudes measured from different plates for the same object
can vary by about $\pm 0.5$ mag.\ (Fig.~\ref{fig:plate_zero_points} below).
Galaxies in the plate overlap regions were therefore
used to match the plates onto a common
magnitude system.

Since the field centres are separated by 5\dg, and the APM scans cover
the full $6\dg \times 6\dg$ area of the UKST plates, 
there is an overlap of about
$6\dg \times 1\dg$ between neighbouring plates.
The 180 eyeballed survey plates give
rise to 493 overlaps, with an average of twelve galaxies
in each overlap.
Only galaxies were used in the overlap comparisons since 
stars brighter than $b_J \approx 16.5$ are very badly saturated on 
UKST J-plates.
These galaxies
are used to calculate a magnitude zero-point offset between
each pair of plates.  An iterative algorithm (Maddox 1988,
\refer{Maddox \etal\ 1990b}) is then used to
find the set of additive plate corrections most consistent with the overlap
zero-point offsets.
Figure~\ref{fig:overlap_mags} plots magnitude difference against mean magnitude
for each matched pair of overlap galaxies (a) before and (b) after plate
matching.
The matching procedure reduces the \rms\ magnitude difference from 0.28 to
0.18, and since there are an average of twelve galaxies per overlap,
the \rms\ error in zero point per overlap
$\approx 0.18/\sqrt{12} \approx 0.05$ mag.
Thus whereas plate matching in the faint survey, with about 3000 galaxies
per overlap, is dominated by systematic errors of order 0.03 mag in the
field corrections, here we are dominated by random errors in
individual galaxy magnitudes.
(The following section suggests that field-effects, which could give
rise to systematic errors, are negligible.)
We do not use the corrections from the faint survey since most bright
galaxies are affected by saturation, and some plates are more saturated
than others.

Figure~\ref{fig:plate_zero_points} is a histogram of plate corrections found
by the matching procedure.
The distribution is roughly Gaussian with $\sigma \approx 0.2$.
Since the eyeballed samples were limited to a fixed APM magnitude limit
before matching, the shallowest fields after matching were up to 0.5 mag
brighter than the mean.
Although many fields were eyeballed to a much fainter limit,
a complete sample could be obtained only by limiting to the matched
magnitude limit of the shallowest field, $m_{mat} = 12.9$.
The number of galaxies to this
limit was only 6250, and so a second pass was made,
eyeballing those survey plates with a matched magnitude limit
shallower than
$m_{mat} = 12.5$ down to this limit.
The plate matching was then repeated
as the larger number of overlap galaxies changes the
field zero-points slightly.
An error in the matching procedure (due to too many iterations)
was subsequently discovered (Loveday 1989), 
resulting in the magnitude limit of the
shallowest field being $m = 12.63$.
After this second pass, 14681 out of 23747 eyeballed galaxies satisfied
the matched magnitude limit $m_{mat} = 12.63$.
An equal-area plot of the final plate zero-point corrections is shown in
Figure~\ref{fig:zero_points_new}.

\section{Internal Tests for Uniformity, Completeness and Classification
Consistency}
\label{sec:int_tests}

\subsection{Field Effects}
\label{sec:field-effects}

In any wide-field telescope such as a Schmidt one expects vignetting towards
the field edges, and indeed this can be seen on the plates.
The APM machine partially corrects
for this by on-line background subtraction.
Maddox \etal\ (1990b) have discussed in detail why vignetting 
is still a problem for faint images.
Basically, vignetting decreases the slope of the measured
density {\em vs.} flux relation so that images measured in vignetted
regions of the plate are analyzed at a higher threshold than those near
the centre.
For brighter images, threshold effects become negligible and for saturated
images vignetting can actually increase the measured magnitude.
This is because saturation density is roughly constant over the whole plate,
\ie independent of vignetting, whereas the sky level {\em is} decreased
by vignetting, and so the sky-subtracted density of saturated images will
tend to increase towards the field edges.

The actual field response function is not necessarily radially symmetric 
due to differential
emulsion desensitization \refer{(Dawe and Metcalfe 1982)}, 
but there are too few
bright galaxies to measure the response as a function of 
2-dimensional position on the plate.
In order to estimate the response function,
we have calculated the galaxy density in annular bins
centred on each Schmidt field and then
summed over all of the survey plates.  By averaging over 180 fields,
any real structure in the galaxy distribution
on a scale $\sim$ plate size should be well averaged out.
To compute the area of each annular bin that lies inside the scanned
region of the plate, 5000 points were thrown down in the central 
$5\dg \times 5 \dg$ square of each plate at
random, avoiding drilled regions.  The galaxy density, normalised by the
density of random points is plotted in Figure~\ref{fig:vig}.
It can be seen that the normalised galaxy density lies
within about one standard deviation of that for a random distribution as far
as the field edges (2.5\dg), although there is marginal evidence for increasing
galaxy density towards the field edges.
Between the field edges and the extreme corners at
$\sqrt{2} \times 2.5\dg \approx 3.54\dg$, the galaxy density drops to
about 80\% of that expected for a random distribution.
For number counts in a Euclidean universe,
$N(m) \propto 10^{0.6(m - m_0)}$,
a twenty percent decrease in counts corresponds to a magnitude
change $\delta m \approx 0.16$ mag.
Only the extreme corners of the field suffer from serious vignetting,
and these observed field effects are small enough not to significantly degrade
the plate matching.

One might expect the observed field response function shown in 
Figure~\ref{fig:vig} to introduce a spurious signal in 
galaxy clustering estimated from the survey.
A powerful technique for studying the effect of systematic errors on
measured galaxy clustering is to compare the angular correlation function \wth\
calculated from pairs of galaxies on the same plate (intra-plate)
and from pairs of galaxies on different plates (inter-plate).
In Figure~\ref{fig:w_ii} we plot intra- and inter-plate estimates of \wth\
for APM-BGC galaxies.
In this and other plots showing \wth, the error bars are determined by
dividing the survey area into four zones and calculating the variance
in \wth\ measured using galaxies in each zone as `centres' for the pair counts.
These error estimates thus include the effects of clustering and
cosmic variance, as well as Poisson statistics.
We see that the intra- (solid symbols) and inter-plate (open symbols)
estimates of \wth\ are in very good agreement, apart from some noise
in the inter-plate estimate on small scales and a small excess in the last
intra-plate estimate.
This last point is less than $2\sigma$ away from the inter-plate estimate
and carries negligible weight in the overall \wth\ from all galaxy pairs
since the ratio of intra- to inter-plate pairs at this separation is
$5\ten{-3}$.

Another test which demonstrates that the field response function has a 
negligible effect on measured clustering is to generate the random
points used in the \wth\ calculation with the {\em same} response function
as the APM-BGC.
In Figure~\ref{fig:w_vig} we plot \wth\ estimated using all (intra- plus
inter-plate) galaxy pairs using both a uniform distribution of random
points (solid symbols) and random points distributed with the same field 
response function as the galaxies (open symbols).
We see no significant difference between the two estimates of \wth,
(indeed, in many cases the symbols are indistinguishable), 
thus confirming that the response function shown in Figure~\ref{fig:vig}
has no deleterious effect on clustering measured from the APM-BGC.

\subsection{Image Classification Consistency and Completeness}

\subsubsection{Consistency in Plate Overlaps}
\label{sec:ovlps}

We have compared the assigned classifications of objects detected on more than
one plate to test the consistency of image classification.
A pair of image detections is included in this
analysis if both image magnitudes are above the magnitude cut-off for their respective
plates and if neither image is in a drilled region.  
%By comparing the classifications of all such
%image pairs a classification consistency table may be 
%produced---Table~\ref{tab:matches}.
Out of 14,358 galaxy images
in the overlaps, where a galaxy is counted on {\em each} plate on which it
was identified, 527
were checked on one plate but not on the other
(\ie the PRE was too `stellar') and 252 objects were 
inspected on both plates, but classified
as a galaxy on one plate, and non-galaxy on the other.  Thus by using the PRE
as a cutoff in deciding which images to eyeball, $3.7\%$ of galaxies 
would not be checked in a single pass.
An additional $1.8\%$ were inspected but mis-classified half of the time.
The actual incompleteness is likely to be
higher than this since a very high surface-brightness galaxy might be classified
as a star on both plates.  Raster scans made of the bright images
in twelve survey fields were used to investigate
the overall completeness of the eyeball survey.  The completenesses inferred
range from 92.6\% to 99.3\% for the 12 fields, with a mean completeness
of 96.3\%, standard deviation 1.9\%.

Table~\ref{tab:matches} shows frequencies of pair classifications
in the overlaps.  The classification codes are defined in
Table~\ref{tab:classns}, and type $-1$ denotes an image that was not checked
(its PRE fell below the cutoff).
Several features about the image classifications emerge from this table:

\begin{enumerate}
 \item Over a third of noise images are also classified as stellar.
       This is because the diffraction spikes and haloes of many bright stars
       are detected as separate images, and it is often difficult to decide
       whether an image should be
       classified as a bright star or noise.
 \item $42\%$ of galaxies classified as elliptical are classified as 
       lenticular on another occasion, and $24\%$
       of lenticulars are also classified as ellipticals.  This illustrates the
       difficulty in distinguishing between the early-type galaxies, and the
       fact that there are more galaxies classified as lenticular than 
       elliptical.
 \item Nearly a quarter of lenticulars are also classified as spirals.
 \item There is a large overlap between irregular galaxies and spirals.
 \item Most galaxies in the unsure type category (type 5) are spirals,
       this is presumably
       because they are the most common.
\end{enumerate}

External checks on the classification reliability and survey completeness
are made by comparison with three other southern galaxy catalogues
in \S~\ref{sec:other_cats}.

\subsubsection{Consistency with Time}
\label{sec:time-dep}

Given that the author had no experience of galaxy classification before
starting this project, one might expect to see systematic variations in
galaxy classification with time.
Most of the fields were originally eyeballed in three batches; the
second pass through was done in two batches.
Table~\ref{tab:trends} shows the number of fields and galaxies eyeballed
in each batch along with the percentage of galaxies of each morphological
classification.

Clearly the ratio of elliptical to lenticular galaxies is much higher 
in batches 1 and 2 than in the later batches.
The total number of early type galaxies (E plus S0) in batch 1 is slightly
higher than the average.
In batch 5 it is much lower than the others, and the fraction of spirals
has risen accordingly.
Nearly all of the galaxies classified as uncertain are in batch 2.
For later batches, a greater effort was made to try and not use this
class where possible.
There are slightly more merged and multiple objects in batch 1 than the
other batches.

To summarize, early type galaxies cannot be reliably distinguished between
elliptical and lenticular.
Batch 5 would appear to be deficient in early type galaxies.

If one combines elliptical and lenticular galaxies together, then the most
discrepant batch regarding balance of morphological types is batch 5,
which includes only 17.3\% early-type galaxies compared with 74.9\% late-types.
In order to investigate what effect this might have on the estimated
correlation function for different morphological types we have divided
the survey into two regions:  those plates included in batch 5 and all
the rest.
We then count galaxy pairs in which both galaxies lie in the same region
(intra-region) and those pairs in different regions (inter-region).
The intra- and inter-region estimates of \wth\ are shown in 
Figure~\ref{fig:w_batch5}.
We see that given the noise in the inter-region estimate, the two estimates
are in reasonable agreement for all types (a), although the intra-region
estimate is biased slightly high relative to the inter-region estimate,
an indication of small but systematic differences in galaxy density
between the two regions.
For early-type galaxies (b), we see a larger systematic difference between
the intra- and inter-region estimates on scales 1--7 degrees, 
$\Delta w \approx 0.05$.
The late-type estimates (c) are in reasonable agreement, although the
inter-region estimate is very noisy on scales less than 2 degrees.

We thus conclude that systematic variations in galaxy classification
reliability may introduce errors in the type-dependent \wth\
of order $\Delta w = 0.05$.

\subsubsection{Consistency with Magnitude}

To illustrate the dependence of classification on magnitude, 
in Figure~\ref{fig:mag_type} we show histograms of the fraction of each 
morphological type as a function of matched $b_J$ magnitude.
The number of ellipticals is roughly constant except for large fluctuations
around $b_J \approx 14$ due to small number statistics.
The number of lenticular galaxies is gradually increasing as the magnitude gets
fainter.
The combined early type number is dominated by the ellipticals at bright
magnitudes and by the lenticulars at fainter magnitudes, where the overall 
trend is for increasing numbers of early type galaxies.
Conversely, the number of spiral galaxies decreases at fainter magnitudes.
This suggests that as the images get fainter and spiral structure is
harder to see, we are increasingly likely to classify a galaxy as a lenticular
rather than a spiral if in doubt.
Note that in the deeper ($b_J = 17.15$) Stromlo-APM Redshift Survey
(Loveday \etal\ 1992) there is a bias {\em against} classifying the fainter
galaxies as early-type, as they are instead classed as `uncertain'.
The fraction of galaxies classified as early type peaks 
near the magnitude limit of the APM-BGC.
The fractions of irregular and `uncertain' galaxies both increase at
fainter magnitudes.

One should be aware of the magnitude-dependent bias in classification 
when interpreting type-dependent clustering results from the APM-BGC.
See \S\ref{sec:clust} for such an analysis.

\section{Completeness and Classification Reliability:
Comparison with other Catalogues}
\label{sec:other_cats}

Two tests of the APM Bright Galaxy Catalogue are made in this section.
Firstly we use the European Southern Observatory (ESO) Survey 
\refer{(Lauberts 1982)}
to identify any galaxies too large to be detected by the APM machine
(see \S\ref{sec:large_ims}) and to check for galaxies found in the
ESO survey which are missed or misclassified in the APM-BGC.
We then compare our galaxy morphological classifications with three other 
southern galaxy catalogues.

\subsection{Identification of Missing Galaxies}
\label{sec:eso}

The ESO Survey, (Lauberts 1982),
is a diameter limited galaxy catalogue, claimed to be complete to one 
arcminute,
but also containing smaller disturbed galaxies, star clusters and
planetary nebulae.
The source material is the ESO(B) Atlas, taken in a blue waveband
similar to the Johnson (B) colour, with the ESO 1m Schmidt in Chile.
This survey covers the southern sky from $-90$ to $-17.5$ degrees in
declination and so is ideal for comparison with the APM Bright Galaxy
Catalogue.

Since the ESO catalogue is diameter limited to 1.0 arcmin,
which corresponds to about 0.9 mm on a Schmidt plate,
{\em all} galaxies too large to be detected by APM
(\S\ref{sec:large_ims}) should be found in the ESO catalogue,
assuming that the ESO and UKST plates reach a similar limiting isophote.
There are an average of 32 ESO galaxies per Schmidt field, with an average
of 29 per field lying within 60 arcseconds of a bright APM image.
Those ESO galaxies outside APM holes
which did not have an APM image identified as a galaxy within 1 arcmin
were inspected by eye.
Table~\ref{tab:eso_unmatched} summarizes the outcome of the ESO-APM
comparison; the final column of this Table indicates the percentage
of ESO galaxies falling into each category.
Those galaxies in category 3 were added to the list of eyeball identifications
(with a flag set to show that they came from the ESO catalogue) and a separate
list of large ESO galaxies (category 2) was compiled (Table~\ref{tab:esobig}).
Thus about 3\% of ESO galaxies were missed or misclassified (mostly
the former) in the APM-BGC, and about 1.5\% are too large to be detected
by APM.

In order to check if any particular sort of galaxies were identified by
ESO but missed in the APM-BGC, apart from the large (category 2) galaxies,
\eg high surface brightness
compact ellipticals, we have looked at the distribution of morphological
types and surface brightness for these galaxies.
Table~\ref{tab:eso_type} shows the numbers and percentages of galaxies
for each morphological type that were identified by ESO but not as galaxies
in the APM-BGC
(`missed', \ie category 3 in Table~\ref{tab:eso_unmatched})
along with galaxies
identified in both the ESO and APM surveys (`found') for comparison.
The relative percentages for the two groups are in reasonable agreement,
except that the missed galaxies contain a larger fraction of
class 8 objects (\ie star-galaxy mergers) and a smaller fraction of
elliptical galaxies, although the numbers of objects in both of these
groups are rather small and subject to large random fluctuations.

The mean surface-brightness of an image, $J_\mu$, measured in
$b_J$ mag arcsec$^{-2}$, is defined by
\be
   J_\mu = b_J + 2.5\log A,
\ee
where $b_J$ is the APM magnitude converted into the $b_J$ system and
$A$ the area above the isophotal threshold in arcsec$^2$.
In Figure~\ref{fig:eso_sb} we plot the surface-brightness frequency histograms
for (a) missed ESO galaxies, (b) found ESO galaxies and (c) all APM galaxies.
In Figure~\ref{fig:eso_sb}(d) we plot the fraction of ESO galaxies that were
missed as a function of surface brightness.
It is clear that the missed ESO galaxies are biased towards high
surface-brightness (HSB; $J_\mu \simlt 22.5$) and in fact this plot provides
a good estimate of our completeness as a function of surface brightness at the
HSB end of the surface brightness range, assuming that the ESO survey
is complete in this regime.
Figure~\ref{fig:eso_sb}(e) shows the fraction of APM galaxies that were
also identified in the ESO survey as a function of surface brightness.
We see that those galaxies common to the two surveys tend to be of lower
surface brightness than typical APM galaxies.
It is not possible to use this comparison to quantify our incompleteness in LSB
galaxies since (i) diameter-limited samples are expected to include
LSB galaxies that would not be included even in a perfect magnitude-limited
sample and (ii) any LSB galaxies that should have been seen in our sample
may have not even been detected by the APM machine and inadvertently
included in category 1 or 2 in Table~\ref{tab:eso_unmatched}.

\subsection{Check on Galaxy Morphological Classification} 
\label{sec:morph_class}

As an external check on the reliability of morphological classifications
of galaxies in the APM-BGC, we have compared morphological classifications
of galaxies common to the APM-BGC and the ESO Surface Photometry Catalogue
(\refer{Lauberts and Valentijn 1989}),
the \refer{Dressler (1980)} cluster catalogue and the
\refer{Corwin \etal\ (1985)} Southern Galaxy Catalogue (SGC).
Since the morphological types in these catalogues
are coded into the de Vaucouleurs
`T-types' (\refer{de Vaucouleurs \etal\ 1976}), (or a nearly equivalent system
for Dressler's data), which sub-divide the coarser 
APM-BGC elliptical-lenticular-spiral classifications, they provide a good check on our
classification reliability.

Tables \ref{tab:corwin}, \ref{tab:dress} and \ref{tab:eso} compare
morphological classifications in the SGC, Dressler and ESO catalogues
respectively with APM-BGC classifications.
The coding is generally in very good agreement except for the overlap between
elliptical and lenticular galaxies, and to a lesser extent between lenticular and spiral
galaxies.
Of course, these catalogues are all shallower than the APM-BGC, and we
expect slightly worse classification reliability at fainter magnitudes.

\section{CCD Calibrations}
\label{sec:photometry}

The purpose of obtaining CCD calibrations was threefold:
Firstly, to determine an overall zero-point for the APM magnitudes;
secondly to check the plate matching procedure by testing for
uniformity of the magnitudes over the
survey area and thirdly to correct the magnitudes for the effects of 
photographic saturation.

\subsection{Observations and Data Reduction}

The photometric observations were carried out with the Siding
Spring Observatory (SSO) 40-inch telescope
CCD system in 1988 and 1990
and with the South African Astronomical
Observatory (SAAO) 40-inch telescope CCD camera in 1988.
Most of the galaxies observed were
chosen from the Stromlo-APM Redshift Survey (\refer{Loveday \etal\ 1992}),
as well as a few sequences
covering a wide magnitude range.
For each galaxy, one 600s exposure was made in $R$ and two 600s
exposures in $B_j$ (SSO) or $B$ (SAAO).
We also observed 21 E-region standard stars from
the list of \refer{Couch and Newell (1980)}.
Flat-field exposures of the
twilight sky were made at the beginning and end of each night,
weather permitting.
Several bias frames were also recorded.
Reduction of the CCD data followed the usual steps of bias-subtraction,
fixing of bad columns and cosmic-rays, and flat-field division.

Images were detected and parameterized using the APM IMAGES software
(\refer{Irwin 1985}).
This software provides the same set of parameters for  CCD images
as are determined by the APM machine from the photographic images,
with the additional options of calculating total image intensities
and deblending merged images.
Both of these options were used.
A detection threshold of 1.5 times \rms\ sky noise was set.
Any contiguous groups of 32 or more pixels ($\approx 10$ arcsec$^2$)
above this threshold were analyzed.
The two blue frames of each galaxy were co-added, after aligning the
detected images, and IMAGES re-run on the co-added frame in order to
minimize readout noise and lower the limiting isophote.

CCD magnitudes were corrected for atmospheric extinction using the
extinction coefficients of Couch and Newell (1980) for the SSO data
and of \refer{Walker (1984)} for the SAAO data.
The standard star observations were then used to transform the CCD
magnitudes into the Couch and Newell (1980)
$B_J$,$R_F$ system using standard colour transform equations
(\eg \refer{Hardie 1962}).
The Couch and Newell system was chosen since their $B_J$ band is
designed to match the UK Schmidt J plate response function.
Data was only used from those nights for which the standard star
(observed $-$ published) residuals were 0.05 mag or less.

\subsection{CCD-APM  Magnitude Comparisons}
\label{sec:ccd}

Determining galaxy magnitudes from the CCD frames and comparison with
the APM magnitude needed to be done interactively due to the
problems of merged images
and occasional break-up of spiral-arm structure into sub-images.

The four sets of CCD images (one red, two blue and co-added blue) were
displayed as ellipse maps and, working from a laser printer grey-scale plot,
any broken images were indicated with the cursor.
A new image intensity and position for the broken images
was obtained by summing the intensities
and calculating an intensity-weighted mean of positions of the bits.
The images in the four frames were then matched up.
The co-added blue frame was used to determine the CCD blue magnitude,
and the difference between the single blue frames and co-added frame was
used to estimate a blue magnitude error.
Comparing the two blue frames is also a good test for photometric
conditions since 
any cloud during the exposure would tend to make the blue magnitudes in
one frame systematically higher or lower than the other.
However, it is of course possible that cloud might affect only the
red frame, which would alter the galaxy colours and hence the
transform to standard magnitudes.
We therefore rejected CCD frames if there was any doubt about
the observing conditions being photometric.
Airmass corrections were applied 
and the CCD magnitudes were transformed into the Couch and Newell
$JF$ system using
transform coefficients determined from the standard star observations.
The CCD and APM images were paired up by displaying ellipse maps of
both and indicating
a few pairs with the cursor.
These pairs were used to fit a 6-constant coordinate transform so that
the remaining image pairs could be found automatically.
Any broken APM
images were summed together in the same way as broken CCD images.
For each galaxy frame a file was written listing natural CCD $J$, $R$
magnitudes; standard $B_J$, $R_F$ magnitudes; APM zero-pointed magnitude and APM
coordinates for each image in the frame.
Bright galaxies were coded according to whether they were isolated images or
merged with a star in the APM and/or CCD image analysis.

In Figure~\ref{fig:galmags} we plot CCD $B_J$ magnitude against matched,
zero-pointed APM magnitude $m_{b_J} = 29.0 - m$.
Merged and multiple galaxies have been excluded from this plot.
The line is a quadratic least-squares fit to the CCD-APM
magnitude relation assuming that the CCD magnitudes are error free.
The transform from $m_{b_J}$ to $b_J$ is given by
\be
 b_J = -4.50 + 1.419 m_{b_J} - 0.00855 m_{b_J}^2.
 \label{eqn:satcor}
\ee
After this saturation correction, the catalogue magnitude limit of 
$m_{b_J} = 16.37$ corresponds to $b_J = 16.44$.
Brightward of $m_{b_J} = 16.37$, the scatter in CCD magnitude about the fit 
is $0.31$ mag.
This compares rather unfavourably with the \rms\ error of 0.16 mag for
the faint APM photometry (Maddox 1988, Maddox \etal\ 1990b), but
is in reasonable agreement with the CCD-COSMOS photometric comparisons
by \refer{Metcalfe \etal\ (1989)} over a similar magnitude range.
A sharp increase in CCD$-$COSMOS magnitude scatter brightward of
$b_J \approx 17$ is very apparent from their Figure~1.
The UK Schmidt J-plates provide reliable images down to $b_J \approx 21$
and for galaxy images brighter than $b_J \approx 16$, saturation
is a serious problem.

Further details of the CCD photometry and the photometric data will be
published separately.

\subsection{Test of matching procedure}

Figure~\ref{fig:ccd-apm_resid_new} plots the CCD$-$APM magnitude residuals in
an equal-area projection.
In the absence of any large-scale gradients in the survey magnitude
calibration, then these residuals should be uncorrelated with both
themselves and with the plate
zero points determined from overlap matching (\S\ref{sec:matching}),
which are plotted in the same projection in Figure~\ref{fig:zero_points_new}.
To test for this we calculated the auto-correlation function of the CCD$-$APM
residuals $r_i$, and the cross-correlation with the plate zero points $z_i$.
The auto-correlation is plotted in Figure~\ref{fig:mag_acor_new}, where
the error bars are given by $\rms(r_i r_j)$ for each $\theta$ bin.
The cross-correlation is plotted in Figure~\ref{fig:mag_xcor_new}, where the
error bars are given by $\rms(r_i z_j)$.
The CCD$-$APM residuals show marginal evidence for being anti-correlated on 
scales $\sim 15\dg$, but only one separation bin shows this anti-correlation
and it is less than a $3\sigma$ effect.
The CCD$-$APM residuals are clearly uncorrelated with the plate zero-points,
confirming the absence of any large-scale gradients in the survey photometry.

Since the \rms\ scatter in individual APM magnitudes
$\approx 0.31$ mag and the \rms\ magnitude error per overlap is only 0.05 mag,
we would require CCD photometry for
$N \simgt (0.31/0.05)^2 \approx 38$ galaxies per field
to improve on the zero-points from
overlap matching.
Therefore the CCD magnitudes were not incorporated into the plate matching
procedure.

\section{The Catalogue}
\label{sec:catalogue}

In compiling the 180 contiguous survey fields into one catalogue,
the Cartesian plate coordinates were converted into right ascension
and declination and only those galaxies within (RA, dec)
boundaries equidistant from field centres
were kept, thus avoiding duplication of galaxies
in the overlaps.
The magnitude limit for each field was set to $b_J = 16.44$
after applying the final
field-corrections from overlap matching (\S\ref{sec:matching})
and correction for saturation (equation~\ref{eqn:satcor}).
Table \ref{tab:summary} summarizes the galaxy numbers in the survey.

Figure~\ref{fig:galdist_all} shows the galaxy distribution in an
equal-area projection for all of the galaxies in the survey
with $b_J \le 16.44$.
Figures~\ref{fig:galdist_early}--\ref{fig:galdist_merge} plot the early-type,
late-type and merged galaxies respectively in the same projection.
The stronger clustering of early-type galaxies over late-type is clearly
visible, and an increase in star-galaxy mergers away from the SGP and
towards the galactic plane is also apparent.
The holes drilled around large images, step wedges and satellite trails
are plotted in Figure~\ref{fig:holes}

Table~\ref{tab:the_cat} presents the data in the APM Bright Galaxy Catalogue
for a single survey field, number 076.  The complete catalogue,
along with the survey field centres and hole positions,
is available from the Astronomical Data Centre 
(http://nssdc.gsfc.nasa.gov/adc/adc.html).
Each column in the table is described below.
\begin{description}

\item[(1) Name:]
Each galaxy name is composed of the survey field number and the $x$, $y$
position of the galaxy on the plate---this should ease location of 
any particular galaxy on the plate material.
The first 3 digits are the SERC field number.
The second set of digits are the $x$-position in millimetres from the
centre of the plate (actually the APM scan centre).
These are preceded by a `+' sign for galaxies to the right (west) of
the plate centre or by a `$-$' sign for galaxies to the left (east) of centre.
The final 3 digits are the $y$ position, again in mm from the plate centre.
A preceding `$-$' indicates galaxies above (north) of the plate centre,
`+' indicates galaxies below (south) of the centre.

\item[(2), (3) RA, dec:]
Right ascension (hours, minutes, seconds) and declination (degrees, arcminutes,
arcseconds) in 1950 coordinates.

\item[(4) $b_J$:]
Matched, saturation corrected $b_J$ magnitude.
Note that although two decimal places of precision are quoted, comparison
with CCD photometry (\S\ref{sec:ccd}) shows that individual galaxy magnitudes
are accurate only to $\approx 0.3$ mag.

\item[(5), (6) maj, min:]
Major and minor diameter in arcseconds at threshold isophote.

\item[(7) p.a.:]
Position angle in degrees measured clockwise from south-north line.

\item[(8) Cl:]
Galaxy classification code, as given in Table~\ref{tab:classns}.
Those galaxies that were only found by cross-checking with the
ESO survey (see \S\ref{sec:eso}) have had 10 added to the classification code.

\end{description}

\section{Type-Dependent Clustering}
\label{sec:clust}

In this section, we calculate the angular and spatial correlation functions for
all unmerged galaxies in the APM-BGC and for early and late type galaxies
separately.
A similar analysis was presented by \refer{Loveday \etal\ (1995)}.
We repeat the analysis here for a number of reasons.
Firstly, use of an earlier, linear correction for photographic saturation from
Loveday (1989) led to a corrected APM-BGC magnitude limit of $b_J = 16.57$
as opposed to $b_J = 16.44$ we obtain from equation (\ref{eqn:satcor}).
This affected the limber inversion carried out by Loveday \etal\ (1995).
Secondly, in generating the random catalogue to correct for the survey
boundaries, we inadvertently generated random points over five additional
fields included in the Stromlo-APM survey, but not in the APM-BGC.
This led to a scale-dependent overestimate of \wth\ by up to 
$\Delta w \approx 0.1$.
Thirdly, in the current analysis we estimate errors by dividing the survey
up into four zones and estimating \wth\ using galaxies in each zone in turn
as centres.
This is likely to give a more realistic estimate of the errors in \wth\ than
the bootstrap resampling technique if the survey is affected by systematic 
variations in morphological classification, as suggested in 
\S\ref{sec:time-dep}.

Following Loveday \etal\ (1995),
we have estimated $w(\theta)$ using the estimator
\be
  w(\theta) = \frac{N_{gg}(\theta) N_{rr}(\theta)}{[N_{gr}(\theta)]^2} - 1
+ \Delta w, \label{eqn:w}
\ee
where $N_{gg}$, $N_{gr}$ and $N_{rr}$ are the number of galaxy-galaxy,
galaxy-random and random-random pairs at angular separation $\theta$ and
$\Delta w$ is a correction for the integral constraint,
\be
  \Delta w = \int\int_{\rm survey} w(\theta_{12}) d\Omega_1 d\Omega_2,
\label{eqn:w_ic}
\ee
(Groth and Peebles 1977).  The correction $\Delta w$ is estimated in
practice by calculating $w(\theta)$ without the correction, integrating
$w(\theta)$ over all elements of solid angle $d\Omega_i$ in the survey area
to obtain $\Delta w$ and recalculating \wth\ with the correction added.  A
stable solution is rapidly reached by iteration.

Figure~\ref{fig:w_type} shows $w(\theta)$ for all, early and late-type 
galaxies in the APM-BGC.  
We have fitted a power law $w(\theta) = A \theta^{1-\gamma}$ from
0.1 to $5\dg$ to these estimates, with results shown in Table~\ref{tab:w}.
We see that early-type galaxies have a slightly steeper power law slope 
and a larger amplitude than
late-type galaxies, in agreement with earlier determinations of the
type-dependent angular correlation function 
(eg. \refer{Davis and Geller (1976)} and  \refer{Giovanelli \etal\ (1986)}).
The integral constraint corrections $\Delta w$
shown in Table~\ref{tab:w} make negligible difference to power-law fits
on scales smaller than $5\dg$ but they do give some idea of possible
systematic errors in the \wth\ estimates on large scales.
The `tail' in the early-type correlation function with $w \approx 0.01$
is almost certainly caused by variations in the classification consistency
discussed in \S\ref{sec:time-dep}.

In order to estimate the spatial correlation functions $\xi(r)$,
we have used these power law solutions in the relativistic version of
Limber's equation (Groth and Peebles 1977, \refer{Phillipps \etal\ 1978})
assuming $q_0 = 0.5$.  The selection function $S(z)$ used in Limber's
equation was determined separately for each galaxy type by smoothing the
observed $N(z)$ for galaxies in the Stromlo-APM survey of the appropriate
type and with $b_J < 16.44$ with a Gaussian of FWHM = 0.01.  
By using the measured selection function $S(z)$ directly in Limber's equation,
we correct to first order for redshift-dependent biases in the morphological
classification.  The resulting
parameters $r_0$ and $B$ for the spatial correlation function $\xi(r) =
(r/r_0)^{-\gamma} = B r^{-\gamma}$ are shown in Table~\ref{tab:w}.  We confirm
that at $r = 1 \hMpc$, the clustering amplitude of early-type galaxies is
more than a factor of three higher than that of late-type galaxies.

\section{Summary}
\label{sec:summary}

We have described the construction of a catalogue of 14,681 galaxies
brighter than $b_J = 16.44$ over a large fraction of the southern sky.
The images were detected and parameterized by scanning 180 United Kingdom
Schmidt Telescope plates with the Automated Plate Measuring system.
Preliminary star-galaxy separation was carried out automatically using
image profiles.
All galaxy candidates were inspected by eye and assigned a morphological
classification.
A completeness of 97\% was aimed for.
By comparing image classifications in overlaps between plates
(\S\ref{sec:ovlps}), we infer an actual completeness of 96.3\%.
As an external check on completeness, we correlated the ESO galaxy catalogue
with the APM-BGC.
We found that about 1.5\% of ESO galaxies are too large to be detected by 
the APM machine (these galaxies are listed in Table~\ref{tab:esobig})
and a further 3\% of ESO galaxies were detected by the APM machine but not
classified as a galaxy in the APM-BGC.
Overall, we estimate that the APM-BGC is at least 95\% complete.
Most of the incompleteness is due either to high surface brightness galaxies
with star-like profiles (Fig.~\ref{fig:eso_sb}(d)) or low surface brightness
galaxies which fall below our detection threshold.
We plan to study surface-brightness selection effects in APM galaxy data in
a future paper.

The reliability of the morphological classification in the APM-BGC was checked
both internally using plate overlaps and externally by comparison with
other catalogues.
We conclude that the APM-BGC does not reliably distinguish between elliptical
and lenticular galaxies; these classes should be combined in any statistical
analysis.
Time-dependent classification effects may produce an error in the 
type-dependent angular correlation function of $\Delta w \approx 0.05$.
We classify fewer galaxies as late type at fainter magnitudes.

The photometry of the survey has been checked using CCD photometry
of 259 galaxies.
We fit a polynomial to the CCD versus APM magnitudes to correct for saturation
and to define the magnitude zero-point.
We find a scatter of 0.31 mag about this relation for individual galaxies.
Comparison of the angular correlation functions calculated
using intra- and inter-plate
pairs of galaxies (\S\ref{sec:field-effects}) shows no evidence for
significant plate-to-plate errors in photometry.
The CCD$-$APM magnitude residuals are uncorrelated with each other
and the plate zero-points, confirming the absence of large scale
gradients in calibration.

The APM Bright Galaxy Catalogue is a reliable, new catalogue of bright 
galaxies which complements
the fainter APM Galaxy Survey and the diameter-selected ESO survey.
The catalogue is about 96\% complete and has essentially zero contamination
since every galaxy has been inspected by eye.
It has proved to be a valuable source catalogue for the Stromlo-APM
Redshift Survey (Loveday \etal\ 1992) and we hope that it will be useful
for other followup work.

\subsection*{Acknowledgments}
The major part of this work was carried out while the author was a 
graduate student at
the University of Cambridge, and I thank the Institute of Astronomy
and Jesus College, Cambridge, for their support.
It is a great pleasure to thank the APM team (Mick Bridgeland,
Pete Bunclark, Mike Irwin and Ed Kibblewhite) for making the APM galaxy 
surveys possible.
Thanks also to Bruce Peterson who obtained some of the CCD photometry,
to Somak Raychaudhury for helpful comments and to Bob Nichol for
generating some finding charts.
I am indebted to George Efstathiou and Steve Maddox for
invaluable guidance, help and discussion throughout this project.

\clearpage

\vspace{.5in}
\centerline{\bf REFERENCES}
\vspace{.2in}
{\refmode
 
Corwin, H.G., de Vaucouleurs, A. and de Vaucouleurs, G., 1985,
Univ of Texas Monographs in Astronomy, No. 4
 
Couch, W.J. and Newell, E.B., 1980, PASP, 92, 746
 
Davis, M. and Geller, M.J., 1976, ApJ, 208, 13
 
Davis, M., Efstathiou, G., Frenk, C.S. and White, S.D.M., 1985,
ApJ, 292, 371
 
Dawe, J.A. and Metcalfe, N., 1982,  Proc.~ASA, 4, 466
 
de Vaucouleurs, G., de Vaucouleurs, A. and Corwin, H.G., 1976,
 Second Reference Catalogue of Bright Galaxies, Univ. Texas Press, Austin
 
Dressler, A., 1980, ApJS, 42, 565
 
Giovanelli, R., Haynes, M.P. and Chincarini, G.L., 1986, ApJ, 300, 77
 
Groth, E.J. and Peebles, P.J.E., 1977, ApJ, 217, 385

Hardie, R.H., 1962, in  Astronomical Techniques, ed. W.A.~Hiltner,
(Univ. Chicago Press), p178
 
Irwin, M.J., 1985, MNRAS, 214, 575
 
Kibblewhite, E.J., Bridgeland, M.T., Bunclark, P. and Irwin, M.J., 1984, in
Astronomical Microdensitometry Conference, NASA Conf. Pub. 2317, p277
 
Lauberts, A., 1982,  The ESO/Uppsala Survey of the ESO(B) Atlas,
European Southern Observatory
 
Lauberts, A. and Valentijn, E.A., 1989,
 The Surface Photometry Catalogue of the ESO-Uppsala Galaxies,
European Southern Observatory
 
Loveday, J., 1989,  Ph.D. Thesis, University of Cambridge
 
Loveday, J., Peterson, B.A., Efstathiou, G. and Maddox, S.J., 1992, ApJ,
390, 338
 
Loveday, J., Maddox, S.J., Efstathiou, G., and Peterson, B.A., 1995,
ApJ, 442, 457
 
Maddox, S.J., 1988,  Ph.D. Thesis, University of Cambridge
 
Maddox, S.J., Sutherland, W.J. Efstathiou, G., and Loveday, J.,
1990a, MNRAS, 243, 692
 
Maddox, S.J., Efstathiou, G. and Sutherland, W.J., 1990b, MNRAS, 246, 433
 
Maddox, S.J., Sutherland, W.J. Efstathiou, G., Loveday, J. and Peterson, B.A.,
1990c, MNRAS, 247, 1P
 
Maddox, S.J., Efstathiou, G., Sutherland, W.J. and Loveday, J.,
1990d, MNRAS, 242, 43P
 
Metcalfe, N., Fong, R., Shanks, T. and Kilkenny, D., 1989, MNRAS, 236, 207
 
Phillipps, S., Fong, R., Ellis, R.S., Fall, S.M. and MacGillivray, H.T.,
1978, MNRAS, 182, 673
 
Raychaudhury, S., 1989, Nature, 342, 251

Raychaudhury, S., Lynden-Bell, D., Scharf, C., and Hudson, M.J., 1994,
Abstracts, 184th AAS Meeting, Minnesota

Shane, C.D. and Wirtanen, C.A., 1967,  Pub. Lick. Obs., 22, Part 1
 
UKSTU handbook, 1983, Royal Observatory, Edinburgh
 
Walker, A.R., 1984,  CCD Photometry using ASPIC Programs, SAAO publication
 
Zwicky, F., Herzog, E., Wild, P., Karpowicz, M. and Kowal, C., 1961--68,
Catalog of Galaxies and Clusters of Galaxies, Vols. I--VI,
California Institute of Technology, Pasadena
}  % end \refmode

\clearpage
\section*{Tables}

\begin{table}[h]
 \begin{center}
 \caption{Image classification coding scheme}
 \label{tab:classns}
 \vspace{0.5cm}
 \begin{tabular}{rl}
  \hline
  \hline
  \multicolumn{1}{c}{Code} & \multicolumn{1}{c}{Description}\\
  \hline
$-1$& Image not checked\\
  0 & Noise (\eg dust, satellite trail, diffraction spike)\\
  1 & Elliptical galaxy\\
  2 & Lenticular galaxy\\
  3 & Spiral galaxy\\
  4 & Irregular/peculiar galaxy or very low surface-brightness\\
  5 & Galaxy, type unsure\\
  6 & Star\\
  7 & Multiple stellar\\
  8 & Merged stellar/galaxy image\\
  9 & Multiple galaxy\\
  \hline
 \end{tabular}
 \end{center}
\end{table}

%\clearpage

\begin{table}[h]
 \begin{center}
  \caption{Comparison of image classifications for objects detected more
	   than once.
	   The classification codes are defined in
	   Table~\protect\ref{tab:classns}.}
  \label{tab:matches}
  \vspace{0.1in}
  \begin{tabular}{rrrrrrrrrrrr}
   \hline
   \hline
 & \multicolumn{11}{c}{Classification Code}\\
       & $-1$   & 0   & 1   & 2   & 3   & 4   & 5   & 6   & 7   & 8   & 9\\
   \hline
 $-1$&  ---&     &     &     &     &     &     &     &     &     &\\
    0&  217&  128&     &     &     &     &     &     &     &     &\\
    1&   53&    7&  314&     &     &     &     &     &     &     &\\
    2&   47&   10&  425&  802&     &     &     &     &     &     &\\
    3&  275&   80&  147&  426& 3292&     &     &     &     &     &\\
    4&   14&    6&    6&   14&  248&  112&     &     &     &     &\\
    5&   16&    0&   23&   16&   85&    6&   12&     &     &     &\\
    6&14376&  303&    5&    4&   13&    1&    0& 4335&     &     &\\
    7& 1507&   50&    0&    2&    5&    1&    0&  611& 1888&     &\\
    8&   83&    8&   13&   29&   44&   14&    0&   67&   42&  256&\\
    9&   11&    5&   12&   21&   23&   16&    1&    0&    2&   17&   79\\
Total&16599&  814& 1005& 1796& 4638&  438&  159&19715& 4108&  573&  187\\
  \hline
  \end{tabular}
 \end{center}
\end{table}

\clearpage
\begin{table}[h]
 \begin{center}
 \caption{The number of fields and galaxies eyeballed in each batch,
	  with the percentage of each morphological classification.}
 \label{tab:trends}
 \vspace{0.1in}
 \begin{tabular}{lrrrrrrrrr}
  \hline
  \hline
 & & & \multicolumn{7}{c}{Galaxy Classification Code}\\
  Batch & Fields & Galaxies & 1 & 2 & 3 & 4 & 5 & 8 & 9\\
  \hline
  1: 24 Mar 87--02 Apr 87 &  18 &  2730 & 35.8 & 3.7 & 47.0 & 1.7 & 0.4 & 8.3 &3.0\\
  2: 29 Apr 87--04 Jun 87 &  25 &  3603 & 26.3 & 5.6 & 49.5 & 3.1 & 9.2 & 4.7 &1.7\\
  3: 12 Jul 87--18 Aug 87 & 130 & 17341 &  5.6 &24.7 & 55.8 & 4.9 & 0.4 & 6.9 &1.8\\
  4: 12 Sep 87--30 Sep 87 &  24 &  1733 &  0.1 &34.4 & 53.2 & 4.3 & 0.0 & 6.8 &1.2\\
  5: 20 Jun 88--22 Jul 88 &  67 &  6926 &  0.3 &17.0 & 71.3 & 3.6 & 0.0 & 5.4 &2.5\\
  \hline
 \end{tabular}
 \end{center}
\end{table}

%\clearpage

\begin{table}[h]
 \begin{center}
 \caption{Results from ESO-APM comparison}
 \label{tab:eso_unmatched}
 \vspace{0.1in}
 \begin{tabular}{cp{4in}c}
  \hline
  \hline
  Category & \multicolumn{1}{c}{Description} & \%\\
  \hline
  1& Estimated to be fainter than the APM-BGC magnitude limit. & 10.7\\
  2& Too large to be detected by APM.  & 1.5\\
  3& Matched up with an APM image not classified as a galaxy. & 3.0\\
  4& Matched up with an APM galaxy. & 78.7\\
  5& No match found; either due to a discrepancy in the coordinate
     systems or a `noise' ESO object. & 6.1\\
  \hline
 \end{tabular}
 \end{center}
\end{table}

\clearpage
\tablecaption{\rm Large ESO galaxies missing from the 
APM Bright Galaxy Catalogue.  RA, dec are in J2000.0 coordinates.}
\tablefirsthead{
   \hline
   \hline
   \multicolumn{3}{c}{\rm RA} & \multicolumn{3}{c}{\rm Dec} & {\rm ESO Name}\\
   \hline}
\tablehead{\multicolumn{7}{l}{\rm Table~\ref{tab:esobig}, continued}\\
   \hline
   \multicolumn{3}{c}{\rm RA} & \multicolumn{3}{c}{\rm Dec} & {\rm ESO Name}\\
   \hline}
\tabletail{\hline}
\begin{tt}
 \begin{supertabular}{rrrrrrc}
  \label{tab:esobig}
   1 &12 &13.5 & -58 &14 &48.1 &113-~G~023 \\
   2 &32 &34.6 & -60 &12 &43.9 &115-IG~018 \\
   3 &13 & 3.8 & -57 &21 &26.1 &116-~G~012 \\
  22 & 8 &33.6 & -57 &26 &34.7 &146-~G~009 \\
  23 &58 &59.8 & -55 &27 &23.4 &149-~G~007 \\
   0 & 1 & 5.8 & -53 &59 &29.8 &149-~GA011 \\
   1 &47 &43.1 & -52 &45 &40.1 &152-~G~024 \\
   3 &35 &57.8 & -52 &39 & 1.7 &156-~G~001 \\
   4 & 4 & 3.0 & -54 & 6 &10.1 &156-~G~036 \\
   4 & 9 &51.3 & -56 & 7 &15.1 &157-~G~003 \\
   4 &15 &45.0 & -55 &35 &31.7 &157-~G~016 \\
   4 &20 & 0.4 & -54 &56 &18.6 &157-~G~020 \\
   4 &21 &27.8 & -55 &56 & 0.3 &157-~G~021 \\
   4 &21 &59.1 & -56 &58 &26.7 &157-~G~022 \\
   4 &27 &37.8 & -55 & 1 &36.5 &157-~G~031 \\
  21 &36 &28.6 & -54 &33 &26.2 &188-~G~012 \\
  23 &33 &16.1 & -54 & 5 &38.3 &192-~G~007 \\
   3 &57 &42.8 & -48 &54 &27.7 &201-~G~012 \\
  21 &19 & 0.3 & -48 &33 &49.8 &236-~G~001 \\
  21 &46 &10.2 & -52 &15 &55.9 &188-IG~019 \\
  21 &52 &42.8 & -48 &15 &15.7 &237-~G~011 \\
  22 & 2 &41.3 & -51 &17 &47.5 &237-~G~027 \\
   0 &39 & 1.3 & -43 & 4 &30.8 &242-IG~022 \\
   3 &38 &44.9 & -44 & 5 &59.1 &249-~G~011 \\
   3 &44 &31.9 & -44 &38 &38.3 &249-~G~016 \\
  21 &15 & 8.0 & -47 &13 &13.2 &286-~G~079 \\
  21 &32 &35.1 & -44 & 4 & 0.1 &287-~G~036 \\
  23 &14 &49.1 & -43 &35 &28.7 &291-~G~010 \\
  23 &18 &23.2 & -42 &21 &41.5 &291-~G~016 \\
  23 &33 &14.7 & -45 & 0 &57.4 &291-~G~029 \\
  23 &33 &17.7 & -45 & 0 &21.4 &291-~G~028 \\
   2 &33 &34.1 & -39 & 2 &45.7 &299-~G~007 \\
   3 &17 &17.5 & -41 & 6 &28.3 &301-~G~002 \\
   3 &51 &40.8 & -38 &27 & 4.0 &302-~G~014 \\
   5 & 7 &42.7 & -37 &30 &44.8 &305-~G~008 \\
  22 &33 &52.5 & -40 &55 &59.8 &345-~G~026 \\
  22 &55 & 0.7 & -39 &39 &41.5 &346-~G~012 \\
  23 & 2 &10.0 & -39 &34 & 8.7 &346-~G~026 \\
  23 &36 &14.5 & -37 &56 &23.3 &347-~G~028 \\
  23 &57 &55.1 & -32 &35 &47.1 &349-~G~012 \\
   1 &22 &43.4 & -36 &34 &44.7 &352-~G~060 \\
   3 &30 &34.8 & -34 &51 &12.3 &358-~G~012 \\
   3 &33 &35.6 & -36 & 8 &23.0 &358-~G~017 \\
   3 &35 &16.7 & -35 &13 &34.6 &358-~G~023 \\
   3 &36 &27.9 & -34 &58 &32.7 &358-~G~028 \\
   3 &37 &11.7 & -35 &44 &41.7 &358-~G~038 \\
   3 &38 &29.0 & -35 &26 &57.9 &358-~G~045 \\
   3 &38 &51.7 & -35 &35 &35.6 &358-~G~046 \\
   3 &42 &19.6 & -35 &23 &36.0 &358-~G~052 \\
   3 &47 & 5.5 & -33 &42 &40.9 &358-~G~065 \\
   4 &12 & 4.3 & -32 &52 &21.0 &359-~G~027 \\
  22 & 8 &28.3 & -34 & 6 &23.2 &404-~G~032 \\
  23 &22 & 4.4 & -33 &22 &31.9 &407-~G~016 \\
   0 &52 &41.7 & -31 &12 &18.6 &411-~G~025 \\
   2 &46 &18.9 & -30 &16 &21.4 &416-~G~020 \\
   3 &18 &15.1 & -27 &36 &35.5 &418-~G~001 \\
   3 &28 &19.0 & -31 & 4 & 4.4 &418-~G~005 \\
   3 &39 &22.6 & -31 &19 &19.9 &418-~G~015 \\
   4 &13 &40.9 & -31 &38 &46.2 &420-~G~012 \\
  22 & 5 &48.7 & -28 &59 &22.4 &467-~G~001 \\
  22 &14 &39.6 & -27 &39 &46.4 &467-~G~028 \\
  22 &27 &22.9 & -31 & 0 &29.1 &467-~G~063 \\
  22 &42 &17.9 & -30 & 3 &28.9 &468-~G~023 \\
  23 &12 & 7.6 & -28 &32 &28.6 &469-~G~019 \\
  23 &27 & 2.2 & -31 &56 &53.3 &470-~G~007 \\
   0 & 9 &53.4 & -24 &57 &11.5 &472-~G~016 \\
   0 &13 &58.7 & -23 &11 & 2.8 &473-~G~001 \\
   0 &47 &34.4 & -25 &17 &32.0 &474-~G~029 \\
   2 &25 & 3.6 & -24 &48 &51.3 &478-~G~028 \\
   2 &26 &21.8 & -24 &17 &32.1 &479-~G~004 \\
   3 & 2 &37.6 & -22 &52 & 3.8 &480-~G~023 \\
   3 &19 &50.8 & -26 & 3 &36.9 &481-~G~020 \\
   3 &39 & 1.8 & -22 &34 &20.0 &482-~G~024 \\
   4 &21 &35.6 & -27 & 7 &40.9 &484-~G~010 \\
   2 &23 & 4.7 & -21 &13 &58.0 &545-~G~011 \\
   3 & 5 &58.7 & -19 &23 &32.5 &547-~G~009 \\
   3 &12 &57.6 & -17 &55 &42.4 &547-~G~020 \\
   3 &24 &38.4 & -19 &18 &22.3 &548-~G~009 \\
   3 &32 & 2.6 & -20 &49 &20.5 &548-~G~031 \\
   3 &44 &50.4 & -21 &55 &20.6 &549-~G~009 \\
   3 &45 &22.0 & -18 &38 & 4.5 &549-~G~012 \\
   3 &48 &14.7 & -21 &28 &27.2 &549-~G~018 \\
   4 & 6 &49.9 & -21 &10 &43.1 &550-~G~007 \\
   4 &42 &14.3 & -20 &26 & 4.1 &551-~G~027 \\
  22 & 8 &19.4 & -19 & 3 &55.8 &601-~G~021 \\
  22 &44 &26.8 & -20 & 2 & 7.3 &603-~G~008 \\
  22 &50 &58.0 & -21 &59 &52.9 &603-~G~019 \\
 \end{supertabular}
\end{tt}

\clearpage
\begin{table}[h]
 \begin{center}
 \caption{Numbers and frequencies of galaxy types missed and found in
	  comparison with the ESO catalogue.
          The classification codes are those
          defined in Table \protect\ref{tab:classns}.}
 \label{tab:eso_type}
 \vspace{0.1in}
 \begin{tabular}{rrrrr}
  \hline
  \hline
  Class & \multicolumn{2}{c}{Missed} &
          \multicolumn{2}{c}{Found}\\
  \cline{2-5}
        & No. & \% & No. & \%\\
  \hline
     1        &6   &3.09   &441   &9.06\\
     2        &20   &10.31   &434   &8.91\\
     3       &136  &70.10  &3404  &69.90\\
     4        &8   &4.12   &320   &6.57\\
     5         &0   &0.00    &34   &0.70\\
     8        &14   &7.22    &89   &1.83\\
     9        &10   &5.15   &148   &3.04\\
 Total       &194 &100.00  &4870 &100.00\\
  \hline
 \end{tabular}
 \end{center}
\end{table}
\clearpage

\begin{table}
 \begin{center}
  \caption{Corwin \etal\ T-type {\em versus} APM eyeball classification frequencies.}
  \label{tab:corwin}
  \vspace{0.1in}
  \begin{tabular}{rlrrrrrrrr}
   \hline
   \hline
   \multicolumn{2}{c}{Corwin} & \multicolumn{8}{c}{APM Eyeball Classification}\\
   T-type & Hubble & 1 & 2 & 3 & 4 & 5 & 8 & 9 & Total\\
   \hline
 $-6$ &cE &     1 &   0 &   0 &   1 &   0 &   0 &   0 &    2 \\
 $-5$ &E0 &    34 &  12 &   1 &   2 &   0 &   1 &   3 &   53 \\
 $-4$ &E+ &    33 &  19 &   0 &   0 &   0 &   2 &   5 &   59 \\
 $-3$ &S0- &   95 &  39 &   4 &   0 &   0 &   1 &   1 &  140 \\
 $-2$ &S0 &    60 &  74 &  52 &   5 &   1 &   0 &   3 &  195 \\
 $-1$ &S0+ &   17 &  21 &  40 &   1 &   2 &   0 &   0 &   81 \\
  0 &S0/a &   2 &   9 &  51 &   2 &   1 &   1 &   0 &   66 \\
  1 &Sa &     0 &   5 &  96 &   5 &   0 &   0 &   0 &  106 \\
  2 &Sab &    1 &   2 &  78 &   0 &   0 &   0 &   0 &   81 \\
  3 &Sb &     3 &   3 & 180 &   0 &   0 &   0 &   1 &  187 \\
  4 &Sbc &    1 &   0 & 154 &   0 &   0 &   0 &   1 &  156 \\
  5 &Sc &     0 &   0 & 193 &   0 &   0 &   3 &   0 &  196 \\
  6 &Scd &    0 &   0 &  80 &   2 &   0 &   0 &   0 &   82 \\
  7 &Sd &     0 &   1 &  55 &  11 &   0 &   0 &   0 &   67 \\
  8 &Sdm &    0 &   0 &  41 &   8 &   0 &   0 &   0 &   49 \\
  9 &Sm &     0 &   1 &  34 &  27 &   0 &   2 &   0 &   64 \\
 10 &Im &     0 &   0 &  19 &  30 &   0 &   2 &   1 &   52 \\
  & Total & 247 & 186 &1078 &  94 &   4 &  12 &  15 & 1636 \\
   \hline
  \end{tabular}
\end{center}
\end{table}
%\clearpage

\begin{table}
 \begin{center}
  \caption{Dressler T-type {\em versus} APM eyeball classification frequencies.}
  \label{tab:dress}
  \vspace{0.1in}
  \begin{tabular}{rlrrrrrrrr}
   \hline
   \hline
   \multicolumn{2}{c}{Dressler} & \multicolumn{8}{c}{APM Eyeball Classification}\\
   T-type & Hubble & 1 & 2 & 3 & 4 & 5 & 8 & 9 & Total\\
   \hline
 $-5$ &E0 &     8 &  14 &   0 &   0 &   0 &   0 &   1 &   23 \\
 $-4$ &E+ &     2 &   1 &   0 &   0 &   0 &   0 &   0 &    3 \\
 $-3$ &S0- &    2 &   4 &   0 &   0 &   0 &   0 &   1 &    7 \\
 $-2$ &S0 &     7 &  19 &  15 &   0 &   0 &   3 &   1 &   45 \\
  0 &S0/a &   0 &   6 &   5 &   0 &   0 &   1 &   0 &   12 \\
  1 &Sa &     1 &   9 &  13 &   0 &   0 &   0 &   1 &   24 \\
  2 &Sab &    0 &   0 &   4 &   0 &   0 &   0 &   1 &    5 \\
  3 &Sb &     2 &   2 &  28 &   0 &   0 &   2 &   0 &   34 \\
  4 &Sbc &    0 &   0 &   3 &   0 &   1 &   1 &   0 &    5 \\
  5 &Sc &     1 &   2 &  12 &   0 &   2 &   0 &   0 &   17 \\
  6 &Scd &    0 &   2 &   6 &   2 &   0 &   1 &   0 &   11 \\
  8 &Sdm &    0 &   0 &   1 &   0 &   0 &   0 &   0 &    1 \\
 10 &Im &     0 &   0 &   1 &   1 &   0 &   1 &   0 &    3 \\
 11 &cI &     2 &   4 &   7 &   0 &   0 &   3 &   0 &   16 \\
 &  Total &  25 &  63 &  95 &   3 &   3 &  12 &   5 &  206 \\
   \hline
  \end{tabular}
\end{center}
\end{table}

\clearpage
\begin{table}
 \begin{center}
  \caption{ESO T-type {\em versus} APM eyeball classification frequencies.}
  \label{tab:eso}
  \vspace{0.1in}
  \begin{tabular}{rlrrrrrrrr}
   \hline
   \hline
   \multicolumn{2}{c}{ESO} & \multicolumn{8}{c}{APM Eyeball Classification}\\
   T-type & Hubble & 1 & 2 & 3 & 4 & 5 & 8 & 9 & Total\\
   \hline
 $-5$ &E0 &    52 &  20 &   8 &   2 &   1 &   0 &   5 &   88 \\
 $-4$ &E+ &    42 &  20 &   1 &   0 &   0 &   3 &   4 &   70 \\
 $-3$ &S0- &  123 &  74 &  23 &   0 &   0 &   3 &   2 &  225 \\
 $-2$ &S0 &    85 & 121 & 121 &   9 &   6 &   8 &   8 &  358 \\
 $-1$ &S0+ &   39 &  50 &  85 &   4 &   0 &   4 &   1 &  183 \\
  0 &S0/a &  21 &  41 & 158 &  11 &   3 &   2 &   4 &  240 \\
  1 &Sa &    21 &  40 & 461 &  22 &   8 &   5 &  11 &  568 \\
  2 &Sab &   13 &  27 & 210 &  10 &   2 &   9 &  19 &  290 \\
  3 &Sb &     6 &   9 & 692 &  24 &   3 &   2 &  26 &  762 \\
  4 &Sbc &    4 &   4 & 304 &  11 &   0 &   3 &  10 &  336 \\
  5 &Sc &     1 &   5 & 287 &  15 &   0 &   9 &  10 &  327 \\
  6 &Scd &    2 &   1 & 476 &  15 &   3 &  13 &  13 &  523 \\
  7 &Sd &     1 &   5 & 155 &  45 &   6 &   6 &  12 &  230 \\
  8 &Sdm &    2 &   0 & 122 &  34 &   0 &   2 &   6 &  166 \\
  9 &Sm &     1 &   1 &  59 &  47 &   0 &   5 &   4 &  117 \\
 10 &Im &     1 &   2 &  56 &  62 &   1 &   5 &   1 &  128 \\
  & Total & 414 & 420 &3218 & 311 &  33 &  79 & 136 & 4611 \\
   \hline
  \end{tabular}
\end{center}
\end{table}

%\clearpage
\begin{table}[htbp]
 \begin{center}
  \caption{Summary of galaxy counts in the complete catalogue
	   ($b_J \le 16.44$).}
  \label{tab:summary}
  \vspace{0.1in}
  \begin{tabular}{rrr}
   \hline
   \hline
   Galaxy Type & Number & \% \\
   \hline
   Elliptical 		&  1791 &  12.2 \\
   lenticular         		&  2648 &  18.0 \\
   Spiral     		&  8217 &  56.0 \\
   Irregular/Peculiar 	&   627 &   4.3 \\
   Unsure     		&   164 &   1.1 \\
   Merged with star 	&   975 &   6.6 \\
   Multiple   		&   259 &   1.8 \\
   \hline
   Total 		& 14681 & 100.0 \\
   \hline
   Holes drilled & 1456 & --- \\
   \hline
  \end{tabular}
 \end{center}
\end{table}

\clearpage
\tablecaption{\rm The APM Bright Galaxy Catalogue for field 076}
\tablefirsthead{
   \hline
   \hline
   \multicolumn{1}{c}{\rm Name} & \multicolumn{3}{c}{\rm RA} & 
   \multicolumn{3}{c}{\rm Dec}
   & $b_J$ & maj & min &{\rm p.a.} & {\rm Cl}\\
   \hline}
\tablehead{\multicolumn{12}{l}{\rm Table~\ref{tab:the_cat}, continued}\\
   \hline
   \multicolumn{1}{c}{\rm Name} & \multicolumn{3}{c}{\rm RA} & 
   \multicolumn{3}{c}{\rm Dec}
   & $b_J$ & maj & min &{\rm p.a.} & {\rm Cl}\\
   \hline}
\tabletail{\hline}
\begin{tt}
 \begin{supertabular}{rrrrrrrrrrrr}
  \label{tab:the_cat}
076-086-127 & 22 &48 &53.59 & -67 &41 &10.0 & 15.65 &  41 &  28 & 150 &   3 \\
076-069-060 & 22 &46 &18.04 & -68 &57 &21.3 & 14.67 &  69 &  30 & 156 &   3 \\
076-060-049 & 22 &44 &39.38 & -69 &10 & 1.4 & 14.28 &  64 &  45 & 177 &   3 \\
076-055-049 & 22 &43 &38.60 & -69 &10 &14.4 & 16.14 &  48 &  20 & 115 &   2 \\
076-039-099 & 22 &39 &49.52 & -68 &15 & 8.3 & 15.88 &  35 &  18 &  55 &   3 \\
076-033-110 & 22 &38 &41.63 & -68 & 2 &49.9 & 15.59 &  43 &  29 &  61 &   2 \\
076-015-079 & 22 &35 & 0.24 & -68 &38 & 8.3 & 16.36 &  35 &  24 & 169 &   2 \\
076-009-052 & 22 &34 & 1.52 & -69 & 7 &57.7 & 15.68 &  36 &  24 & 175 &  13 \\
076+022-124 & 22 &27 &41.23 & -67 &47 &50.9 & 16.17 &  31 &  21 &  11 &   3 \\
076+058-111 & 22 &20 &30.09 & -68 & 0 &35.6 & 16.41 &  33 &  21 &   8 &   2 \\
076+065-133 & 22 &19 &12.69 & -67 &35 &25.1 & 16.44 &  24 &  17 &   5 &   3 \\
076+069-114 & 22 &18 &17.69 & -67 &56 &23.4 & 16.10 &  34 &  23 &  10 &   3 \\
076+100-103 & 22 &12 & 6.23 & -68 & 6 &33.0 & 16.18 &  51 &  30 & 124 &   4 \\
076+105-125 & 22 &11 &27.23 & -67 &41 &33.8 & 15.52 &  43 &  31 & 108 &   3 \\
076+106-066 & 22 &10 &13.79 & -68 &46 &43.0 & 15.97 &  51 &  25 &  99 &   3 \\
076+112-059 & 22 & 8 &42.68 & -68 &54 &31.7 & 14.56 &  62 &  43 &  72 &   2 \\
076+117-126 & 22 & 9 & 4.61 & -67 &39 &45.6 & 15.64 &  39 &  33 & 171 &   2 \\
076+121-120 & 22 & 8 &14.48 & -67 &45 &21.6 & 16.26 &  46 &  14 & 156 &   3 \\
076-116+007 & 22 &57 &29.69 & -70 & 7 &17.7 & 16.11 &  31 &  22 & 150 &   8 \\
076-113-015 & 22 &56 &20.68 & -69 &43 & 6.7 & 16.42 &  41 &  17 &  52 &   3 \\
076-108-037 & 22 &54 &51.23 & -69 &19 &14.2 & 14.15 &  91 &  39 &  31 &   3 \\
076-099-007 & 22 &53 &32.00 & -69 &53 &20.9 & 15.89 &  43 &  24 &  80 &   3 \\
076-090+043 & 22 &52 &28.98 & -70 &50 &27.6 & 14.82 &  48 &  46 & 132 &   2 \\
076-087+046 & 22 &51 &56.43 & -70 &53 &53.9 & 15.26 &  75 &  26 & 174 &   3 \\
076-078-031 & 22 &48 &34.08 & -69 &29 & 8.8 & 15.68 &  34 &  27 & 155 &   3 \\
076-060-032 & 22 &44 &43.85 & -69 &28 &37.5 & 15.40 &  59 &  29 & 165 &   3 \\
076-058+021 & 22 &44 &58.11 & -70 &27 &38.5 & 15.57 &  70 &  19 &  11 &   3 \\
076-055-040 & 22 &43 &39.90 & -69 &20 &13.8 & 16.15 &  56 &  14 & 124 &   3 \\
076-044-006 & 22 &41 &31.72 & -69 &58 &52.4 & 16.07 &  32 &  28 &  81 &   3 \\
076-026+041 & 22 &37 &56.04 & -70 &52 &18.7 & 16.11 &  31 &  23 &  18 &   9 \\
076-024-008 & 22 &37 &10.65 & -69 &57 &20.6 & 15.75 &  46 &  21 &  40 &   8 \\
076+001+023 & 22 &31 &43.20 & -70 &32 &25.9 & 14.69 &  60 &  45 &  41 &   3 \\
076+003-009 & 22 &31 &25.89 & -69 &56 & 0.7 & 16.26 &  28 &  28 &  90 &   2 \\
076+006-014 & 22 &30 &42.23 & -69 &50 &58.5 & 16.17 &  41 &  17 & 107 &   8 \\
076+020+043 & 22 &27 &32.88 & -70 &54 &33.2 & 16.22 &  84 &   9 &  31 &   3 \\
076+037+029 & 22 &23 &41.67 & -70 &38 &37.1 & 15.03 &  48 &  39 &  65 &   3 \\
076+038-044 & 22 &24 & 5.04 & -69 &16 &45.3 & 16.42 &  33 &  24 &  45 &   2 \\
076+039+036 & 22 &23 &15.83 & -70 &46 &13.9 & 16.35 &  65 &   9 &  86 &   3 \\
076+044+039 & 22 &22 & 3.71 & -70 &49 &19.9 & 16.39 &  33 &  20 & 161 &   3 \\
076+047-006 & 22 &21 &54.39 & -69 &58 &21.2 & 16.28 &  33 &  27 & 107 &   3 \\
076+056+021 & 22 &19 &29.76 & -70 &27 &48.5 & 16.43 &  29 &  14 &  14 &   3 \\
076+058+029 & 22 &19 & 0.17 & -70 &37 & 7.8 & 15.76 &  36 &  34 & 115 &   2 \\
076+070+034 & 22 &16 &14.59 & -70 &42 & 0.2 & 16.06 &  38 &  24 &  69 &   3 \\
076+097+053 & 22 & 9 &46.39 & -71 & 0 &52.9 & 16.28 &  26 &  25 & 117 &   3 \\
076+099+045 & 22 & 9 &34.21 & -70 &51 & 7.6 & 15.37 &  43 &  26 &  80 &   8 \\
076+099-022 & 22 &10 &48.25 & -69 &36 &55.1 & 15.04 & 104 &  17 & 106 &   3 \\
076+100+022 & 22 & 9 &39.87 & -70 &25 &37.6 & 16.41 &  36 &  19 & 169 &   4 \\
076+111+021 & 22 & 7 &21.16 & -70 &23 &54.8 & 15.58 &  63 &  14 &  21 &   3 \\
076+110+033 & 22 & 7 &16.35 & -70 &37 & 2.2 & 16.21 &  42 &  14 & 102 &   8 \\
076-099+103 & 22 &55 &56.05 & -71 &56 &36.4 & 16.12 &  32 &  17 & 148 &   3 \\
076-058+109 & 22 &46 & 4.85 & -72 & 5 &52.3 & 16.11 &  49 &  15 &  84 &   3 \\
076-053+090 & 22 &44 &36.86 & -71 &45 &56.8 & 16.42 &  37 &  21 &  77 &   2 \\
076-050+077 & 22 &43 &54.56 & -71 &31 &10.3 & 16.12 &  68 &  10 &   9 &   3 \\
076-049+120 & 22 &44 & 0.94 & -72 &18 &59.0 & 16.06 &  33 &  32 &  57 &   3 \\
076-046+073 & 22 &42 &46.51 & -71 &27 & 2.5 & 16.01 &  45 &  32 &  70 &   3 \\
076-044+080 & 22 &42 &29.15 & -71 &35 &10.7 & 15.32 &  45 &  35 &  50 &   2 \\
076-030+087 & 22 &39 &17.03 & -71 &43 &23.9 & 15.86 &  34 &  27 &  33 &   3 \\
076-020+075 & 22 &36 &42.77 & -71 &30 & 8.5 & 15.28 &  50 &  26 &  62 &   3 \\
076-017+057 & 22 &35 &54.92 & -71 &10 &20.8 & 16.29 &  46 &  17 &  86 &   3 \\
076+021+079 & 22 &27 & 4.40 & -71 &34 &35.9 & 16.34 &  54 &  12 & 103 &   3 \\
076+024+086 & 22 &26 &28.53 & -71 &42 & 4.2 & 15.89 &  35 &  32 & 172 &   1 \\
076+027+083 & 22 &25 &38.44 & -71 &38 &35.9 & 16.41 &  30 &  26 &  86 &   2 \\
076+030+066 & 22 &24 &59.38 & -71 &19 &47.6 & 16.27 &  32 &  30 &  88 &   3 \\
076+033+060 & 22 &24 &31.23 & -71 &12 &33.6 & 16.34 &  31 &  19 & 170 &   4 \\
076+038+095 & 22 &22 &54.57 & -71 &52 & 7.0 & 15.85 &  59 &  16 &  91 &   3 \\
076+060+089 & 22 &17 &45.33 & -71 &43 &57.2 & 16.43 &  34 &  20 &  23 &   3 \\
076+060+074 & 22 &17 &57.49 & -71 &27 &30.9 & 14.93 &  76 &  25 & 165 &   8 \\
076+082+120 & 22 &12 & 4.38 & -72 &16 &27.0 & 16.06 &  63 &  11 & 105 &   3 \\
076+083+092 & 22 &12 &13.65 & -71 &45 &15.4 & 14.81 &  54 &  39 & 166 &   3 \\
076+092+096 & 22 &10 & 8.55 & -71 &49 &27.6 & 15.79 &  60 &  16 & 172 &   3 \\
076+095+106 & 22 & 9 & 3.54 & -71 &59 &38.4 & 15.57 &  44 &  35 & 122 &   4 \\
076+100+072 & 22 & 8 &43.43 & -71 &21 & 3.7 & 15.22 &  59 &  21 & 117 &   3 \\
076+106+104 & 22 & 6 &35.49 & -71 &57 & 3.6 & 15.96 &  59 &  15 & 179 &   3 \\
076+108+086 & 22 & 6 &23.34 & -71 &35 &58.7 & 15.64 &  58 &  29 &  55 &   3 \\
 \end{supertabular}
\end{tt}

\begin{table}[htbp]
 \begin{center} \caption{Angular correlation function results for all,
		early and late type galaxies in the APM-BGC.} 
\vspace{0.5cm} 
\label{tab:w} 
\begin{math} 
\begin{array}{lccccc} 
\hline
\hline 
{\rm Type} & \gamma & A & \Delta w & B & r_0\\ 
\hline 
All & 1.85 \pm 0.14 & 0.18 \pm 0.03 & 2.9\ten{-3} & 16.4 \pm 1.5 & 4.5 \pm 0.6\\ 
Early & 1.93 \pm 0.16 & 0.35 \pm 0.05 & 4.6\ten{-3} & 40.6 \pm 3.5 & 6.8 \pm 0.8\\
Late & 1.79 \pm 0.16 & 0.15 \pm 0.03 & 1.6\ten{-3} & 11.3 \pm 1.0 & 3.9 \pm 0.5\\ 
\hline 
\hline 
\end{array} 
\end{math} 
\vspace{0.2cm}

Note.---Power-law fits ($ w = A
\theta^{1-\gamma}$) were made over the range 0.1--$5\dg$. The integral
constraint $\Delta w $ is estimated from the observed $w(\theta)$.  The
amplitude, $B$, and corresponding scale length, $r_0$ are for the spatial
correlation function inferred from inverting Limber's equation.
\end{center}
\end{table}

\clearpage
\section*{Figure Captions}

\begin{description}

\refstepcounter{figure}\label{fig:fields}
\item[Figure \thefigure] 
An equal-area projection of the fields included in the APM Bright Galaxy
Catalogue.
The UKSTU field numbers are indicated, along with lines of constant RA and dec.

\refstepcounter{figure}\label{fig:profile}
\item[Figure \thefigure] 
Areal profiles for (a) stars and (b) galaxies in the 
APM magnitude range 12.5--12.6 on one survey plate.

\refstepcounter{figure}\label{fig:pre}
\item[Figure \thefigure] 
Profile residual error $\veps$ plotted against APM magnitude
for one survey field.
Dots denote stars, plus signs galaxies, asterisks multiple-stars,
six-pointed stars star-galaxy mergers and open crosses multiple galaxies.
Open squares show `noise' images.

\refstepcounter{figure}\label{fig:overlap_mags}
\item[Figure \thefigure] 
Overlap magnitude errors (a) before and (b) after plate matching.

\refstepcounter{figure}\label{fig:plate_zero_points}
\item[Figure \thefigure] 
Histogram of plate zero-point corrections from matching procedure.

\refstepcounter{figure}\label{fig:zero_points_new}
\item[Figure \thefigure] 
Plate corrections after final matching, 
plotted in the same  equal area projection as Figure~\ref{fig:fields}.
The line to the bottom-left shows a magnitude correction
of 0.5 mag.

\refstepcounter{figure}\label{fig:vig}
\item[Figure \thefigure] 
Galaxy density as a function of distance 
from the field centre, obtained by stacking all of the survey plates
and normalizing by a random distribution of points within the
field boundaries and outside drilled regions.

\refstepcounter{figure}\label{fig:w_ii}
\item[Figure \thefigure] 
The angular correlation function \wth\ for APM-BGC galaxies from
intra-plate pairs (solid symbols) and inter-plate pairs (open symbols).
Error bars are determined from the variance between four zones.

\refstepcounter{figure}\label{fig:w_vig}
\item[Figure \thefigure] 
The angular correlation function \wth\ from all pairs of APM-BGC galaxies 
measured using a uniform random catalogue (solid symbols) and a random
catalogue with the field response function shown in 
Figure~\protect{\ref{fig:vig}} (open symbols).

\refstepcounter{figure}\label{fig:w_batch5}
\item[Figure \thefigure] 
The angular correlation function \wth\ from
intra-region pairs (solid symbols) and inter-region pairs (open symbols)
for (a) all galaxy types, (b) early type galaxies and (c) late type galaxies.
The two regions consist of (i) plates eyeballed in batch 5 
(Table~\protect{\ref{tab:trends}}) and (ii) all the rest.

\refstepcounter{figure}\label{fig:mag_type}
\item[Figure \thefigure] 
Histograms of fraction of each morphological type
as a function of matched $b_J$ magnitude.

\refstepcounter{figure}\label{fig:eso_sb}
\item[Figure \thefigure] 
Surface-brightness frequency histograms
for (a) missed ESO galaxies, (b) found ESO galaxies, (c) all
APM galaxies, (d) fraction of galaxies missed and 
(e) fraction of APM galaxies also detected by ESO survey.

\refstepcounter{figure}\label{fig:galmags}
\item[Figure \thefigure] 
Total CCD $B_J$ magnitudes plotted against APM magnitudes for
259 unmerged galaxies.
The line shows the quadratic fit that we have used to convert
photographic magnitudes to the $B_J$ system.
The scatter about this line brightward of $m_{b_J} = 16.37$
(shown by the vertical dotted line) is 0.31 mag.

\refstepcounter{figure}\label{fig:ccd-apm_resid_new}
\item[Figure \thefigure] 
CCD$-$APM magnitude residuals
plotted in the same  equal area projection as Figure~\ref{fig:fields}.
The line in the bottom-left corner shows a magnitude residual
of 0.5 mag.

\refstepcounter{figure}\label{fig:mag_acor_new}
\item[Figure \thefigure] 
Auto-correlation function of the CCD$-$APM magnitude residuals.

\refstepcounter{figure}\label{fig:mag_xcor_new}
\item[Figure \thefigure] 
Cross-correlation of plate zero points with
CCD$-$APM residuals.

\refstepcounter{figure}\label{fig:galdist_all}
\item[Figure \thefigure] 
All 14,681 galaxies in the APM bright galaxy survey 
plotted in the same equal area projection as Figure~\ref{fig:fields}.

\refstepcounter{figure}\label{fig:galdist_early}
\item[Figure \thefigure] 
The 4439 early-type (elliptical plus lenticular) galaxies.

\refstepcounter{figure}\label{fig:galdist_late}
\item[Figure \thefigure] 
The 8844 late-type (spiral plus irregular) galaxies.

\refstepcounter{figure}\label{fig:galdist_merge}
\item[Figure \thefigure] 
The 975 star-galaxy and galaxy-galaxy merged images.

\refstepcounter{figure}\label{fig:holes}
\item[Figure \thefigure] 
The 1456 holes drilled in the survey.

\refstepcounter{figure}\label{fig:w_type}
\item[Figure \thefigure] The angular correlation functions $w(\theta)$ 
	for all (open circles), early-type (filled circles) and late-type
	(filled squares) galaxies in the APM Bright Galaxy Catalogue.
	The dotted lines show power-law fits from 0.1 to $5\dg$.

\end{description}

% Finally, include the figures
%\epsfverbosetrue
\epsfxsize=0.95\textwidth
\setcounter{figure}{0}

% bright:plot_fields.f; bright:fields.lis
\begin{figure}[p]
\epsfbox{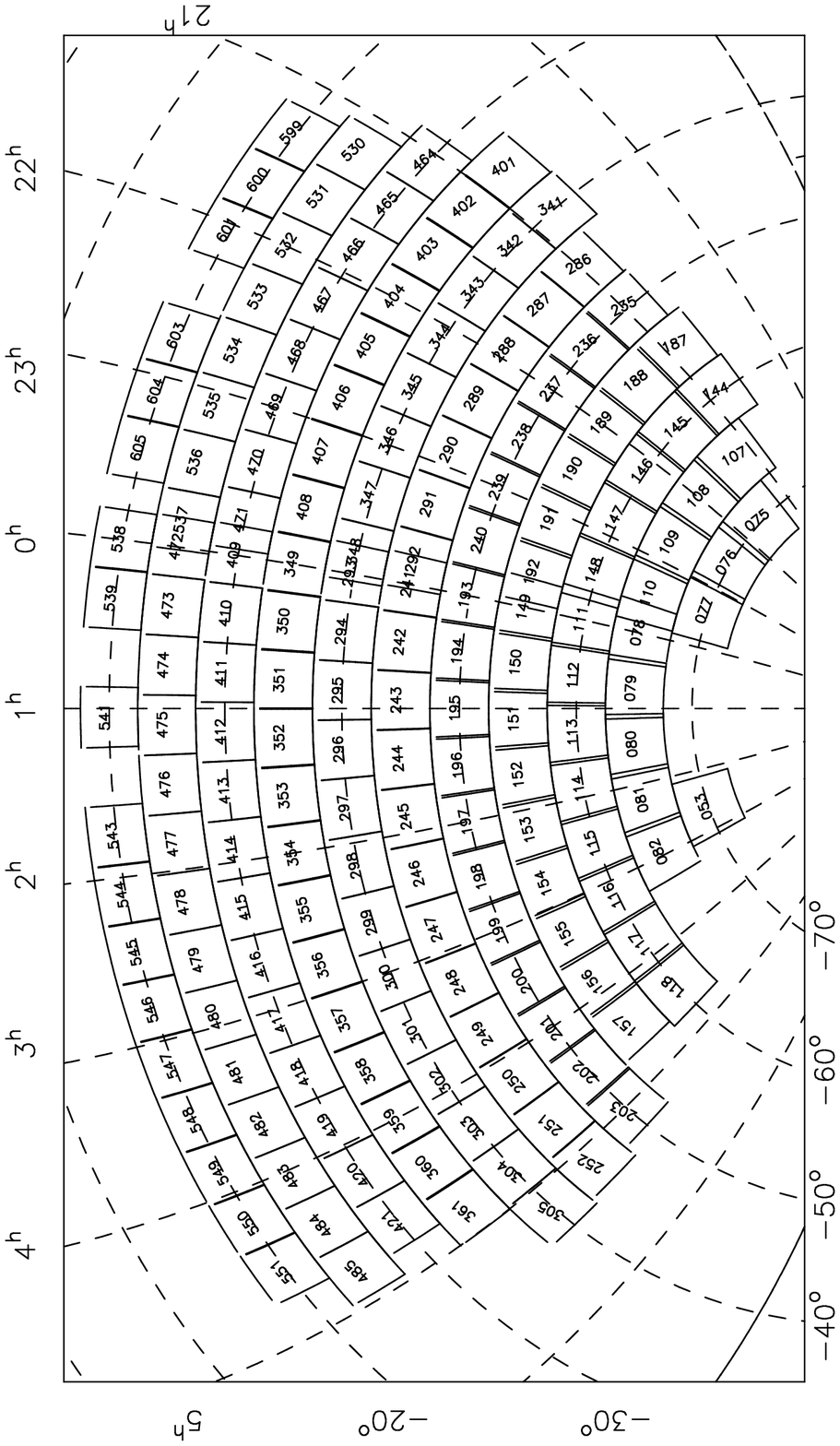}
\caption{\relax}
\end{figure}

% bright:profile_plot.f; survey:f080.cls; mag 12.5-12.6 10% of stars
\begin{figure}[p]
\epsfbox{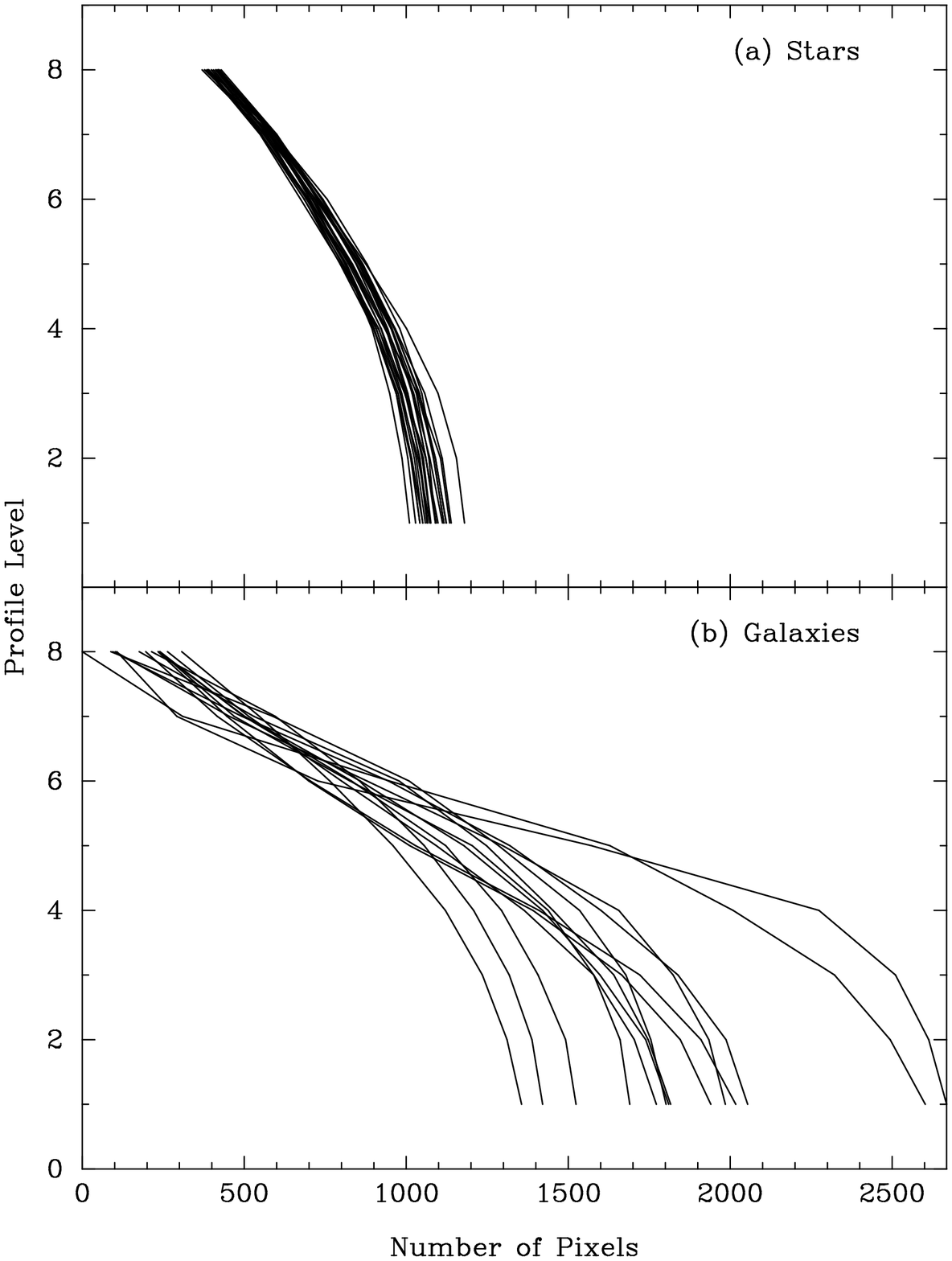}
\caption{\relax}
\end{figure}

% bright:eyeball.f; survey:f405.eye, j6231.ptr, j6231.rcl, j6231.cls
\begin{figure}[p]
\epsfbox{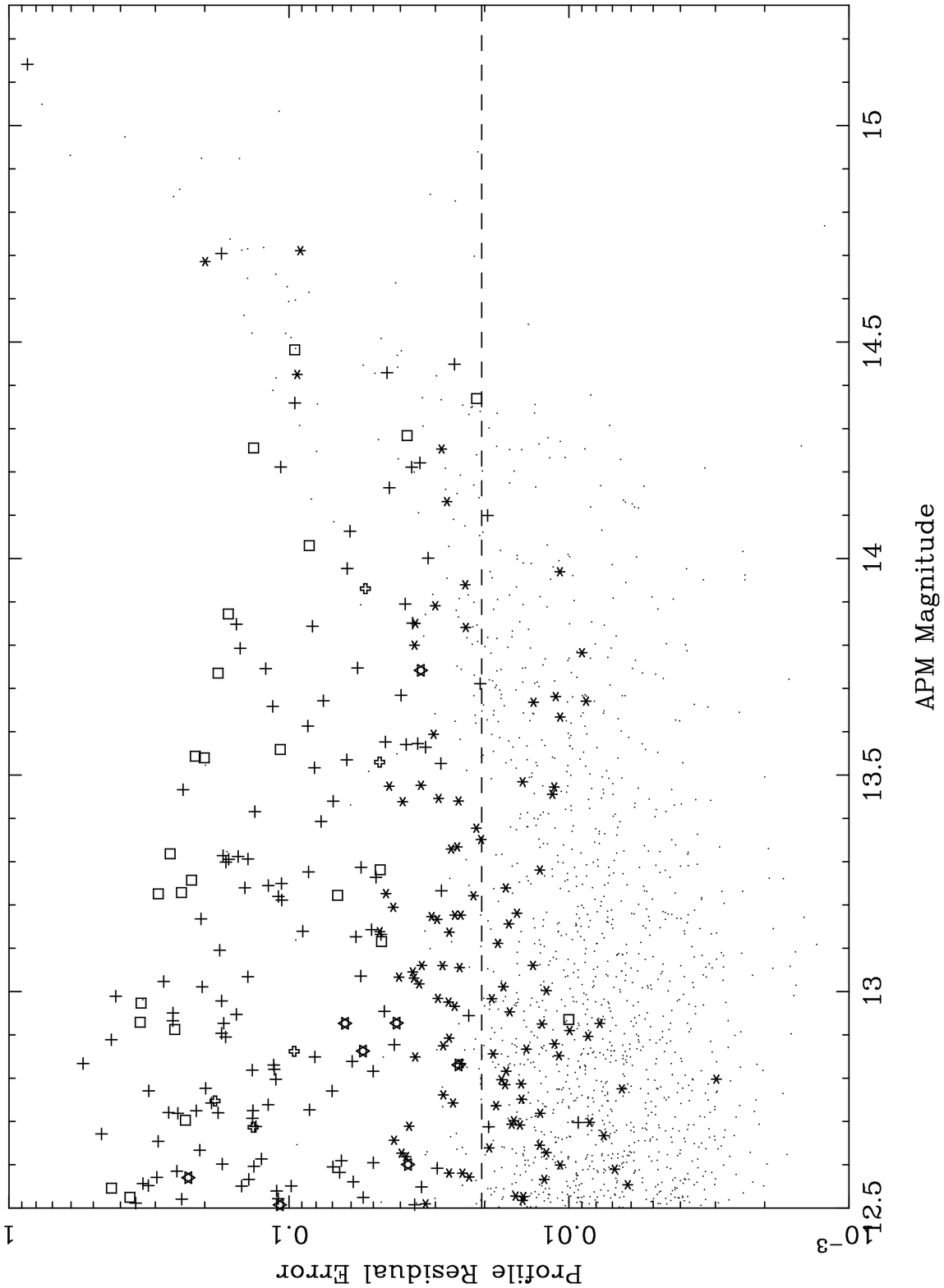}
\caption{\relax}
\end{figure}

% bright:ovlps_plot.f; bright:ovlps_plot.dat
\begin{figure}[p]
\epsfbox{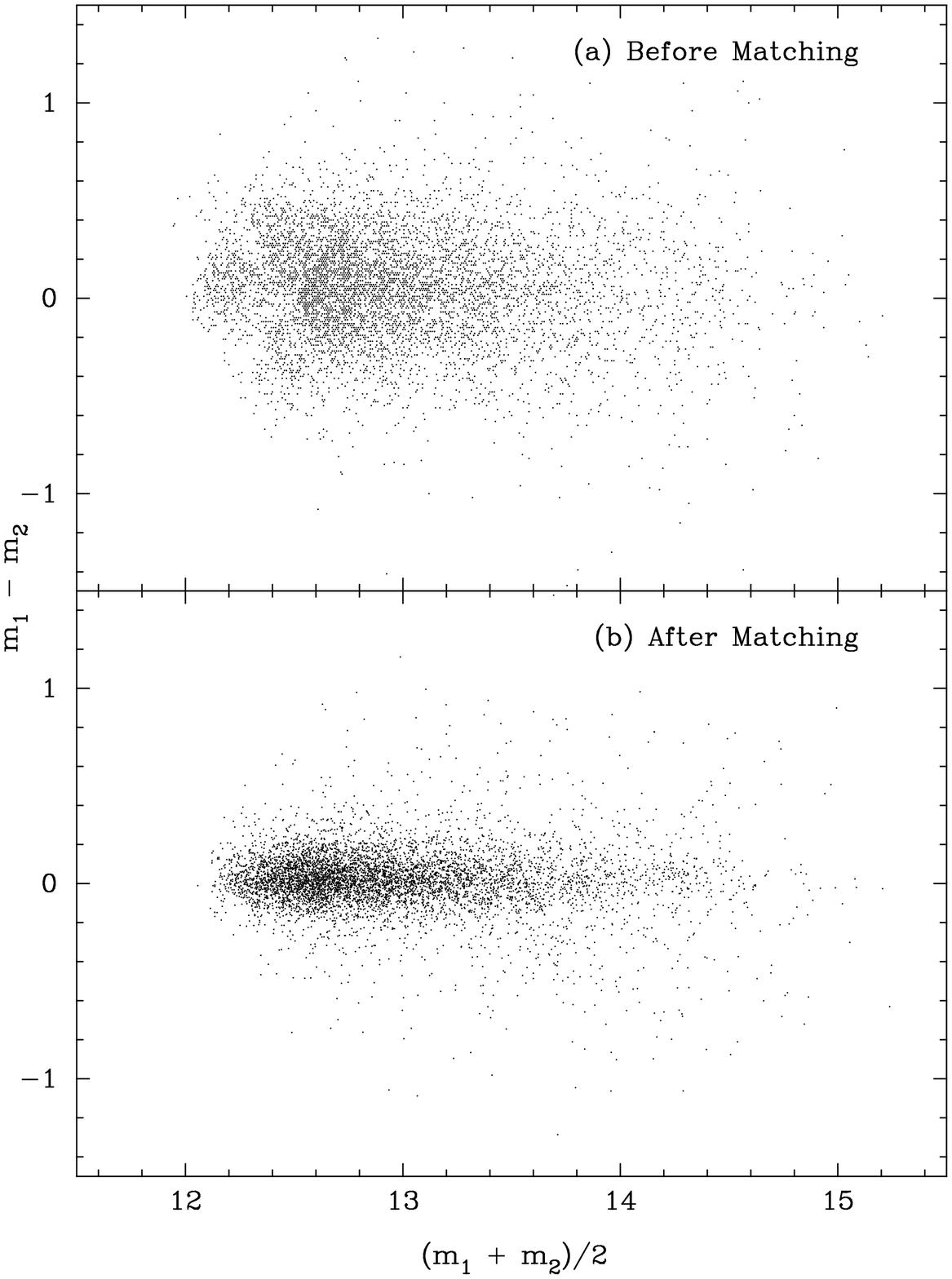}
\caption{\relax}
\end{figure}

% bright:zerohist.f; bright:tran100.dat
\begin{figure}[p]
\epsfbox{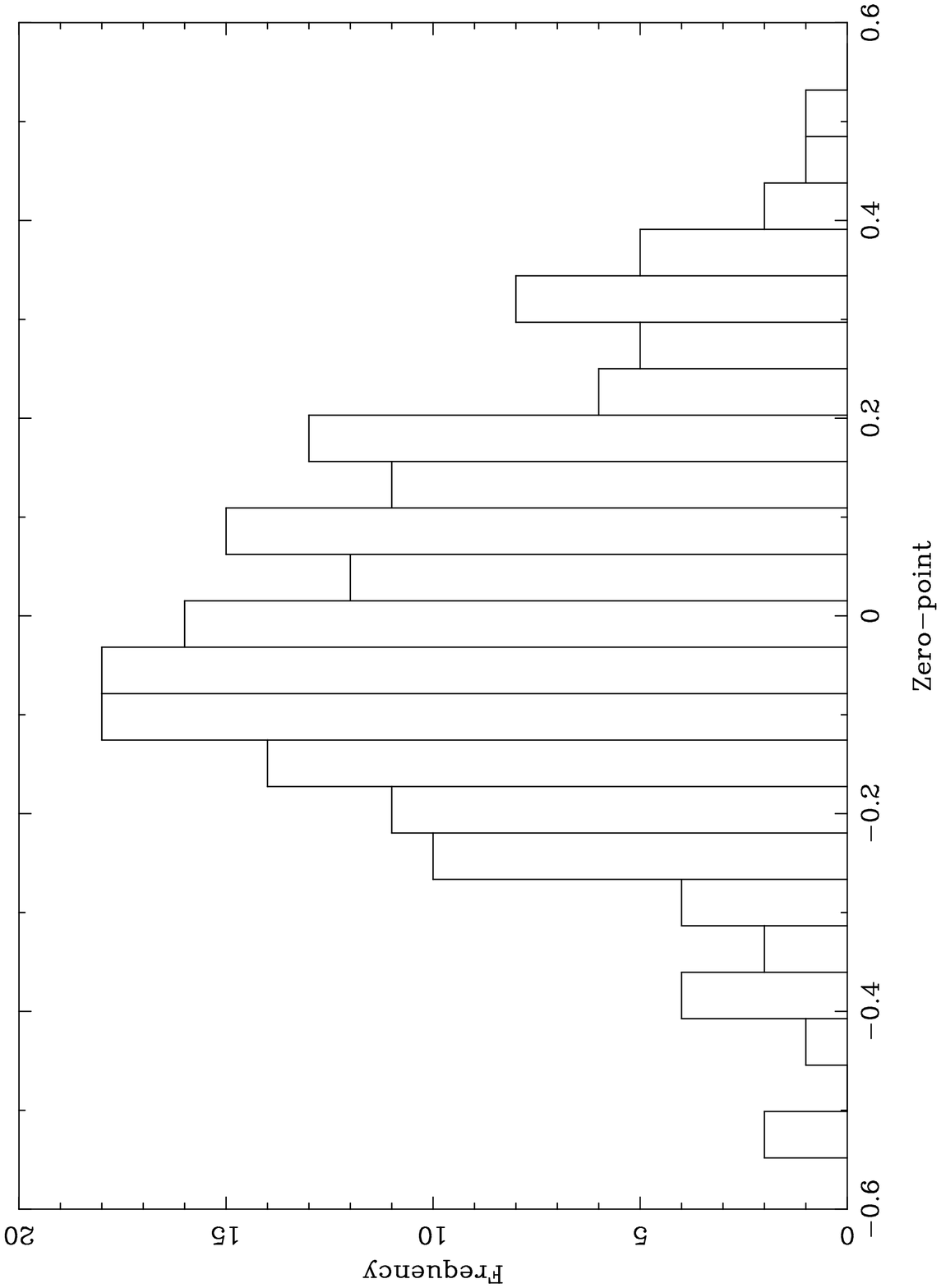}
\caption{\relax}
\end{figure}

% bright:magplot.f; bright:tran7.dat
\begin{figure}[p]
\epsfbox{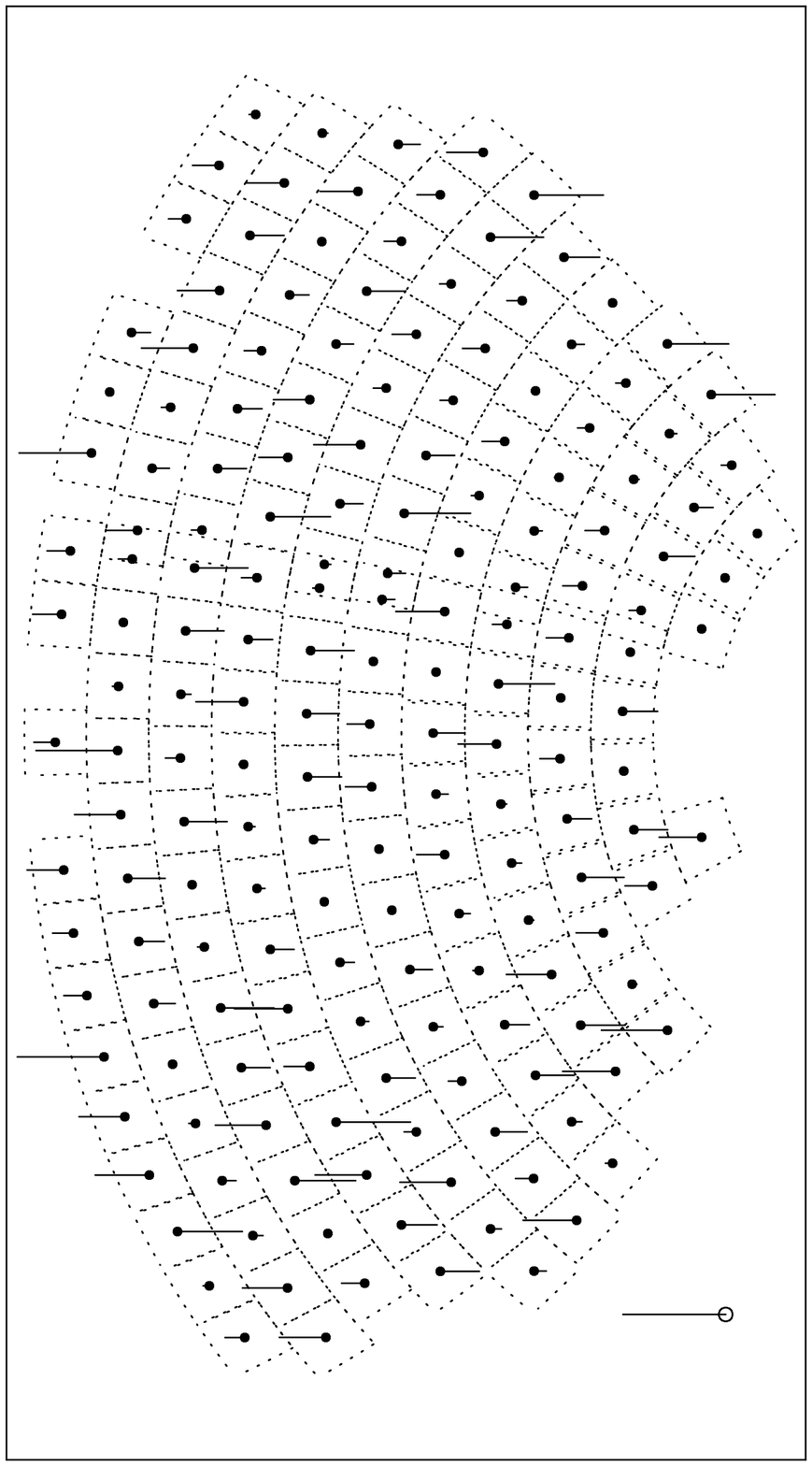}
\caption{\relax}
\end{figure}

% bright:vignette.f; bright:mags_7.lis, $survey/all/*.all; galxs only
\begin{figure}[p]
\epsfbox{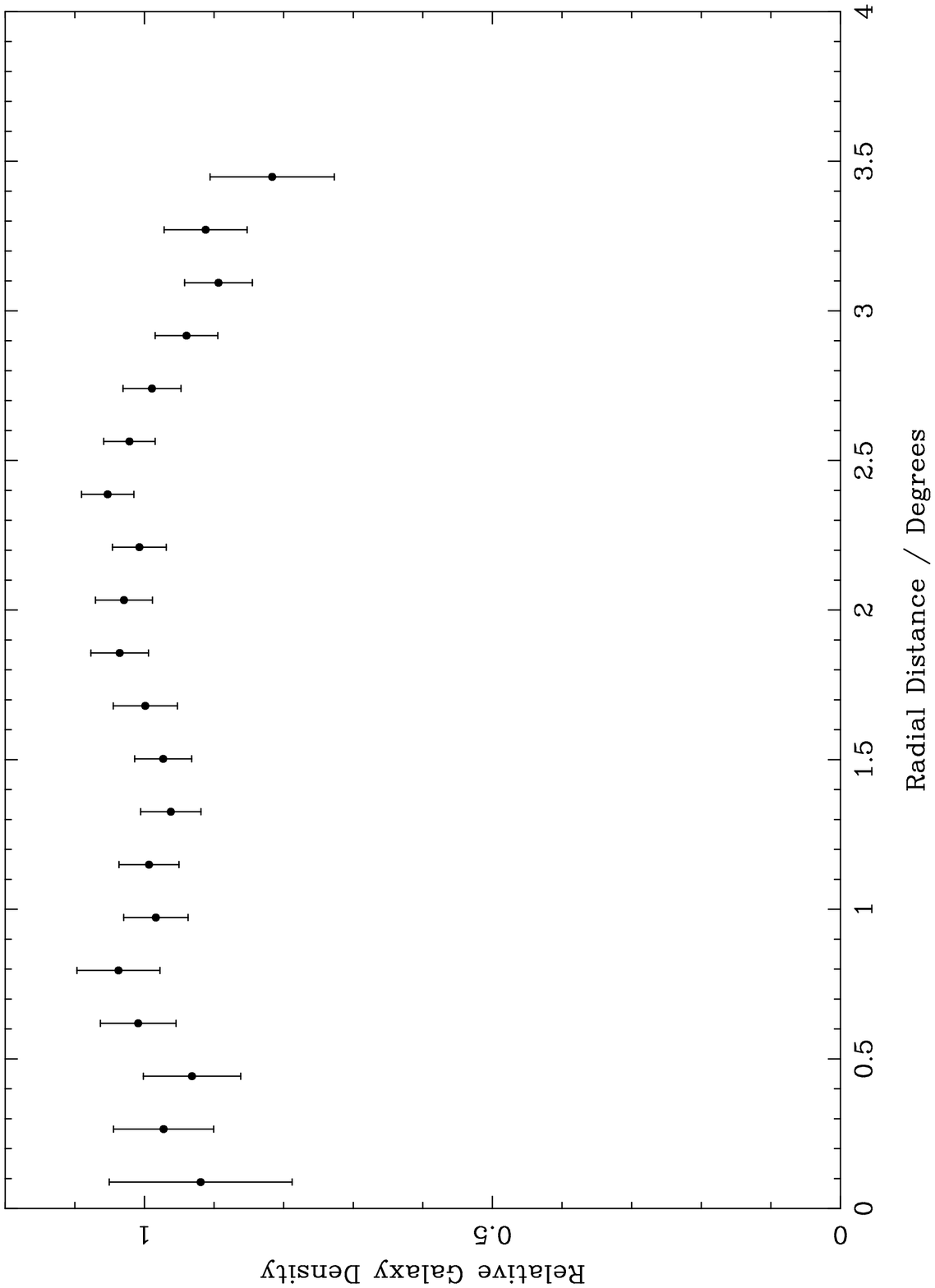}
\caption{\relax}
\end{figure}

% corr/xip2.f; cordat:wp_all_intra.dat and wp_all_inter.dat
\begin{figure}[p]
\epsfbox{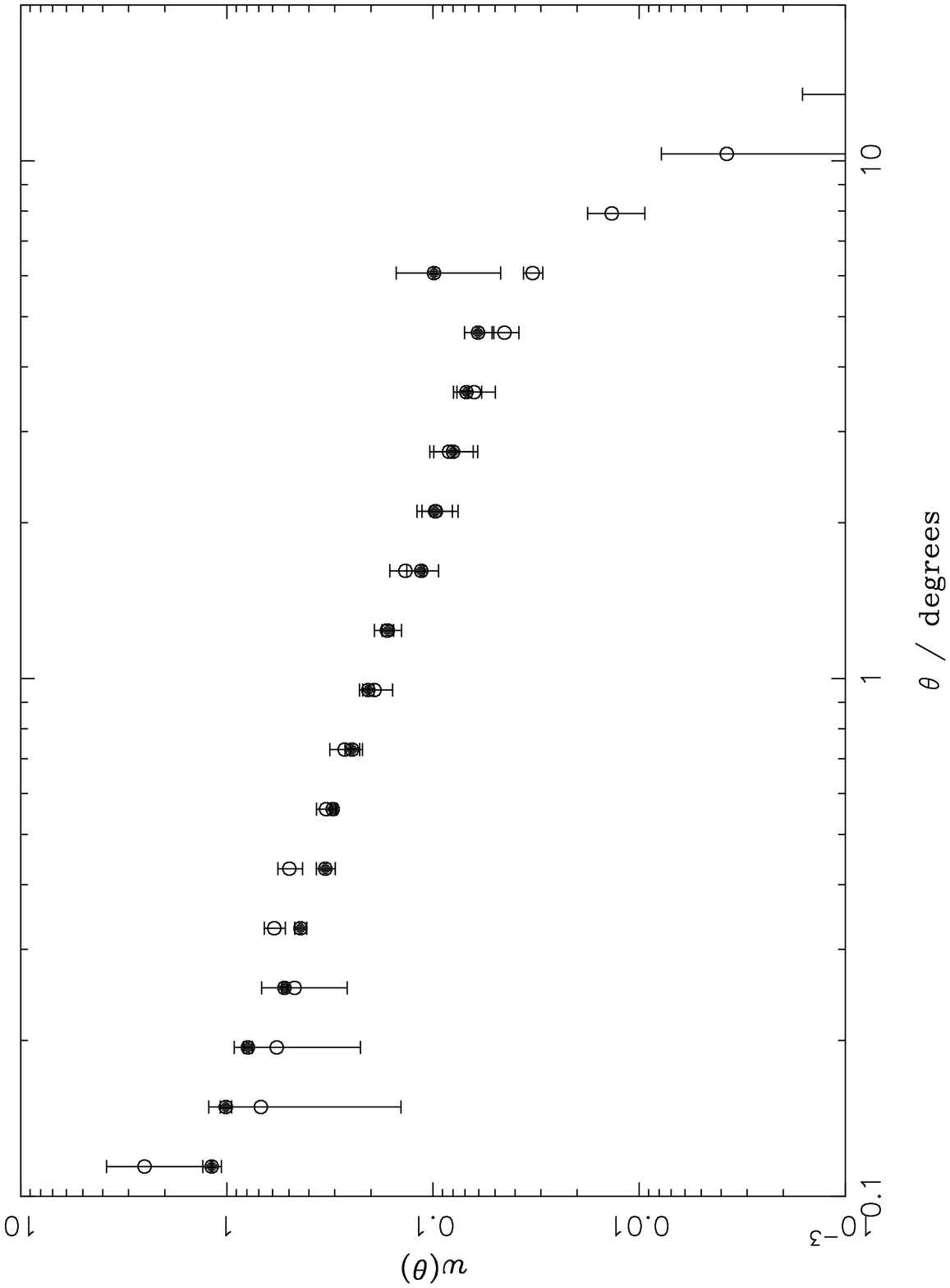}
\caption{\relax}
\end{figure}

% corr/xip2.f; cordat:wp_all_z_n2.dat and wp_all_vig_z_n2.dat
\begin{figure}[p]
\epsfbox{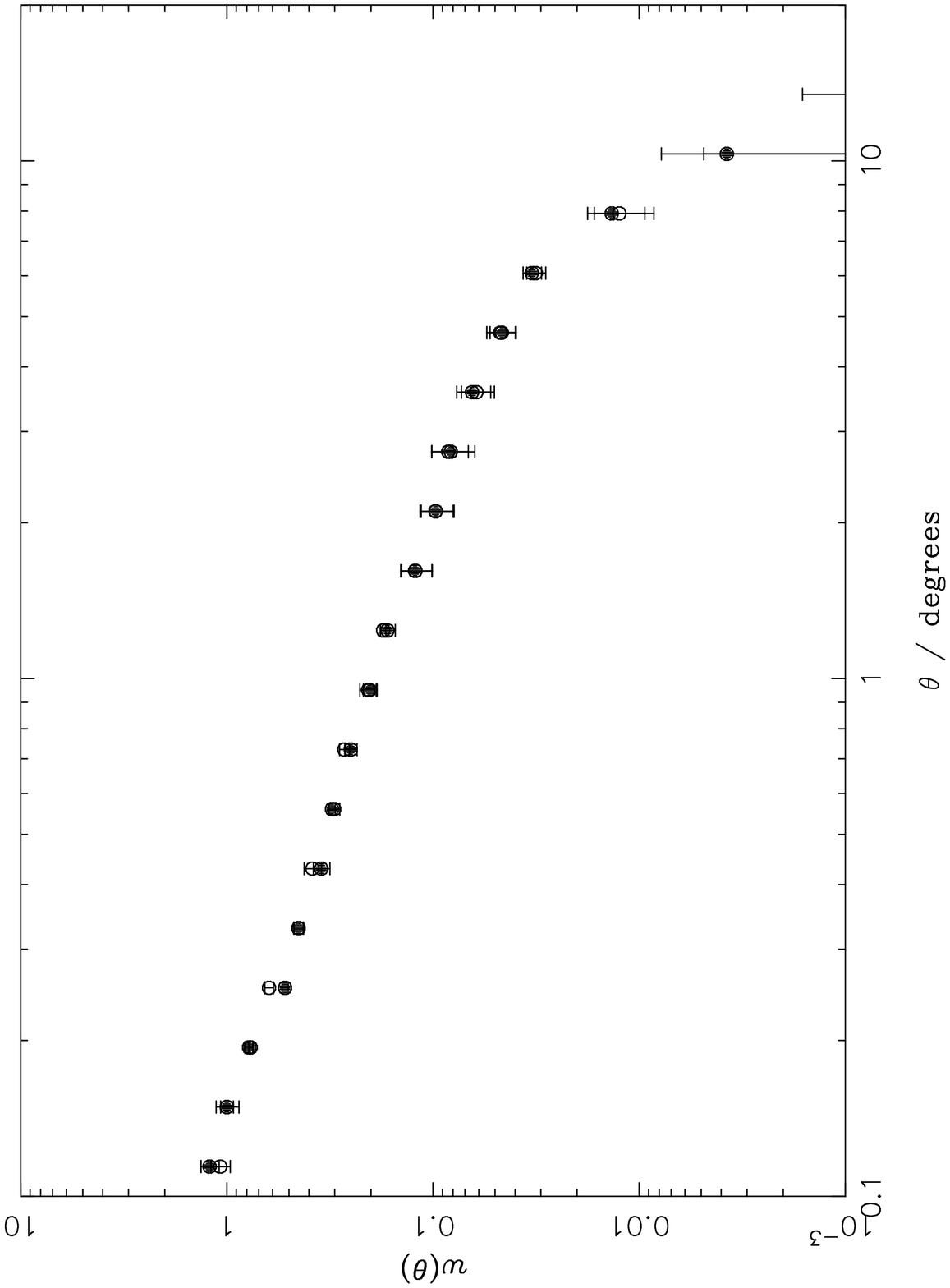}
\caption{\relax}
\end{figure}

% corr/xip_mult.f; (a) cordat:wp_all_b5intra.dat + wp_all_b5inter.dat;
% (b) wp_early_b5intra.dat + wp_early_b5inter.dat; 
% (c) wp_late_b5intra.dat + wp_late_b5inter.dat
\begin{figure}[p]
\epsfbox{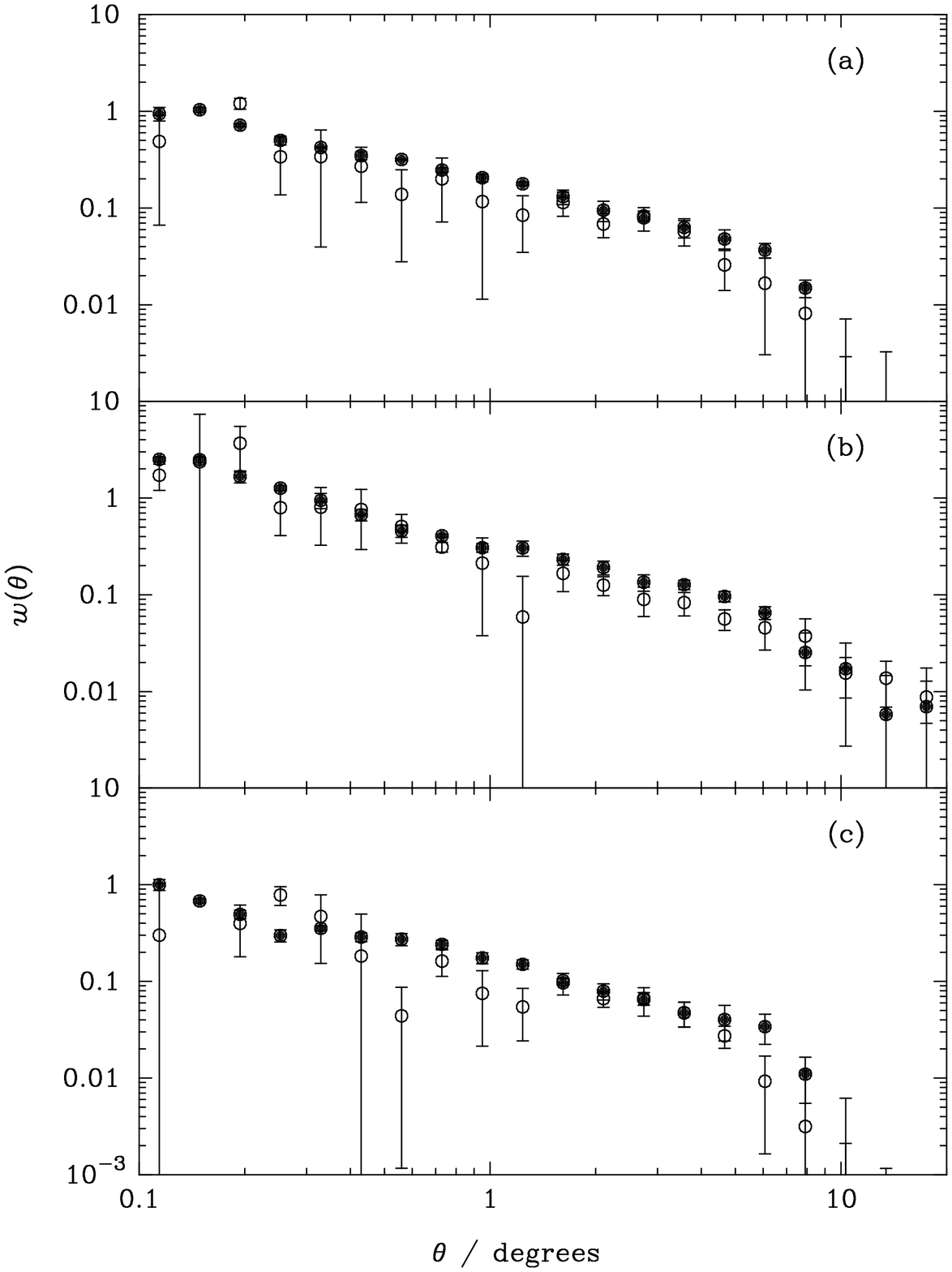}
\caption{\relax}
\end{figure}

% bright:mag_type.f; bright:master7.dat
\begin{figure}[p]
\epsfbox{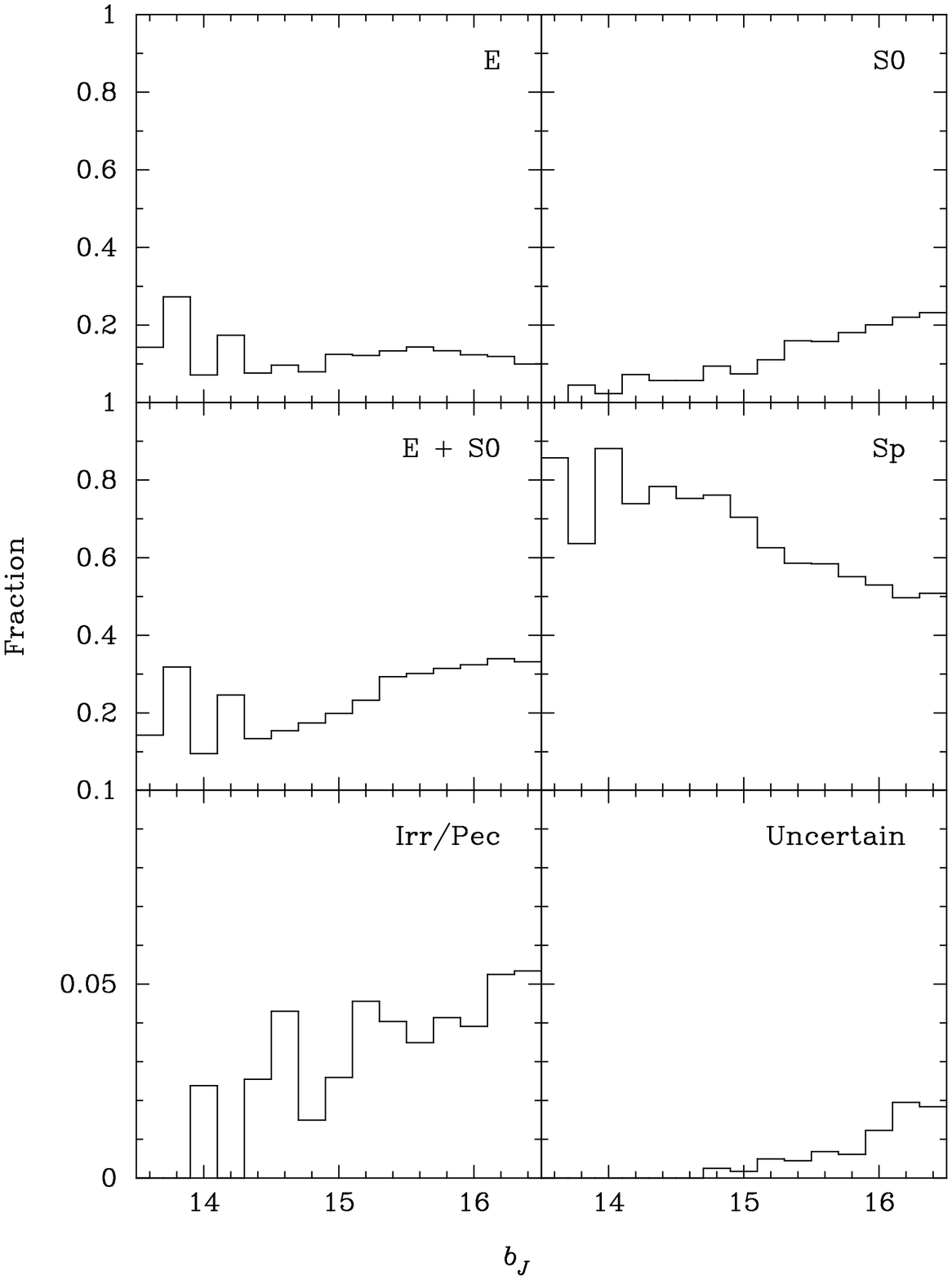}
\caption{\relax}
\end{figure}

% bright:eso_sb.f; survey:apmes.lis, master4d.lis
\begin{figure}[p]
\epsfbox{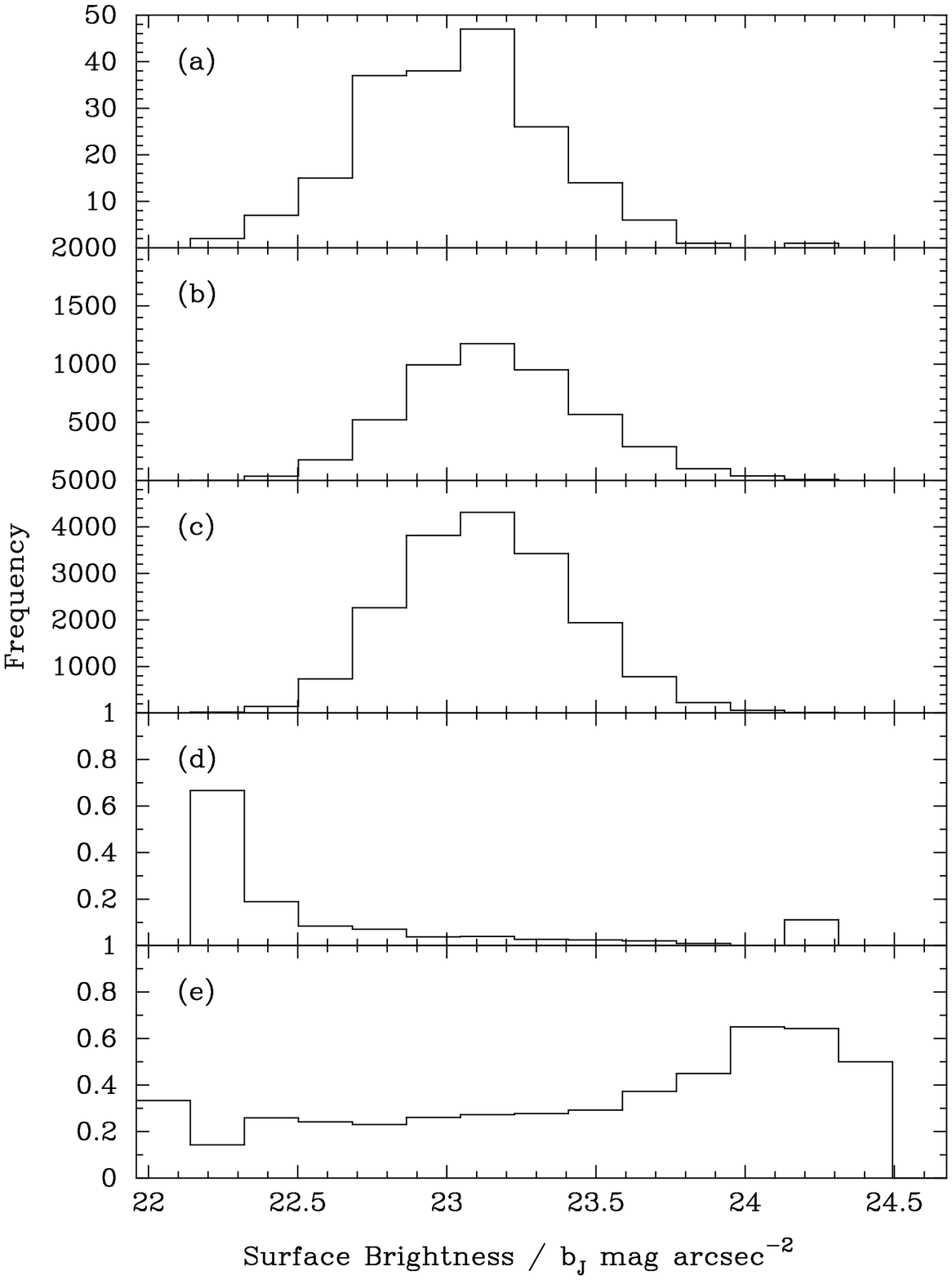}
\caption{\relax}
\end{figure}

% ccd:satcal.f; ccd:ccd.seq, ccd88.seq
\begin{figure}[p]
\epsfbox{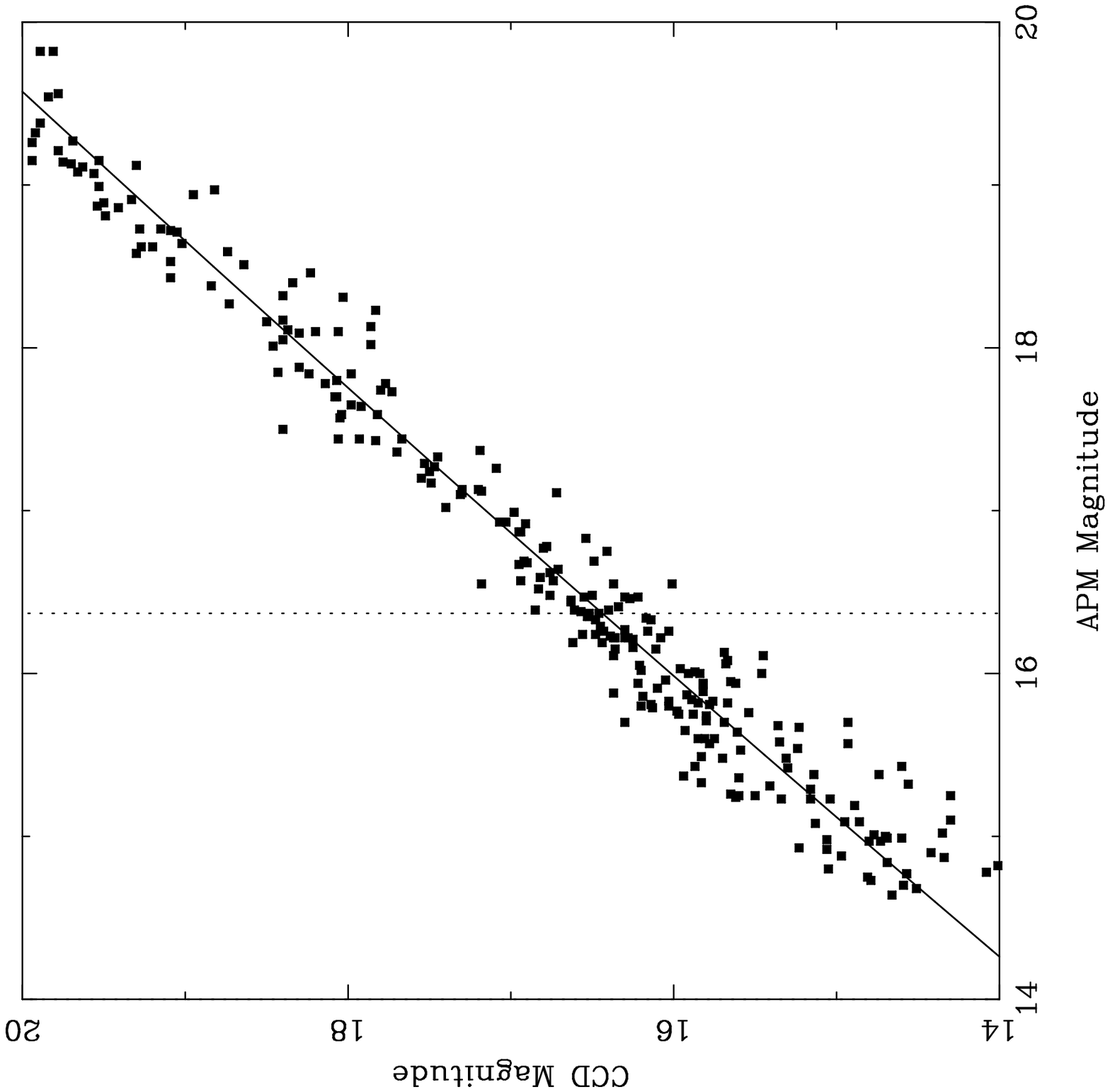}
\caption{\relax}
\end{figure}

% bright:magplot.f; ccd:satcal.lis
\begin{figure}[p]
\epsfbox{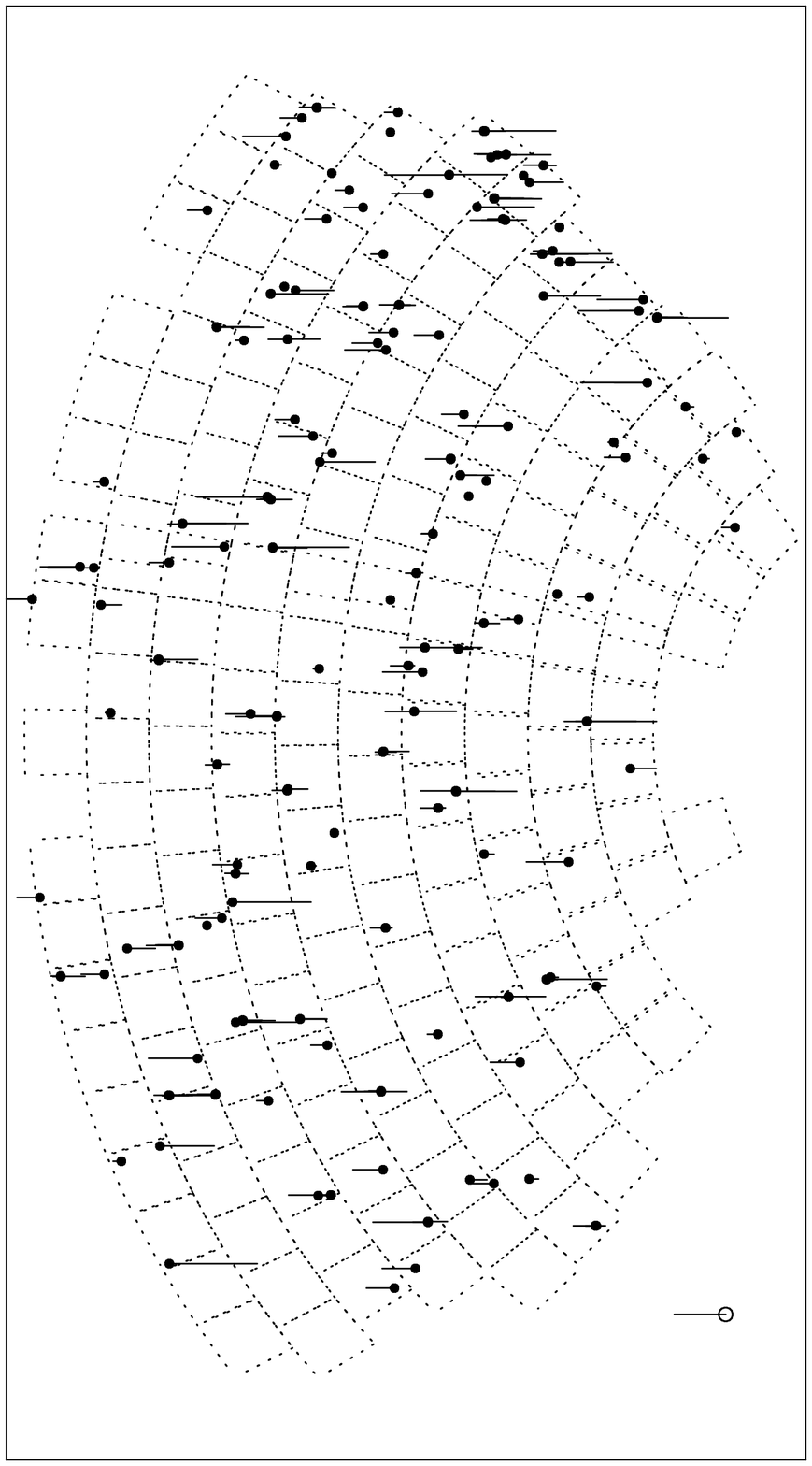}
\caption{\relax}
\end{figure}

% bright:magcor.f; ccd:satcal.lis (auto-correlation)
\begin{figure}[p]
\epsfbox{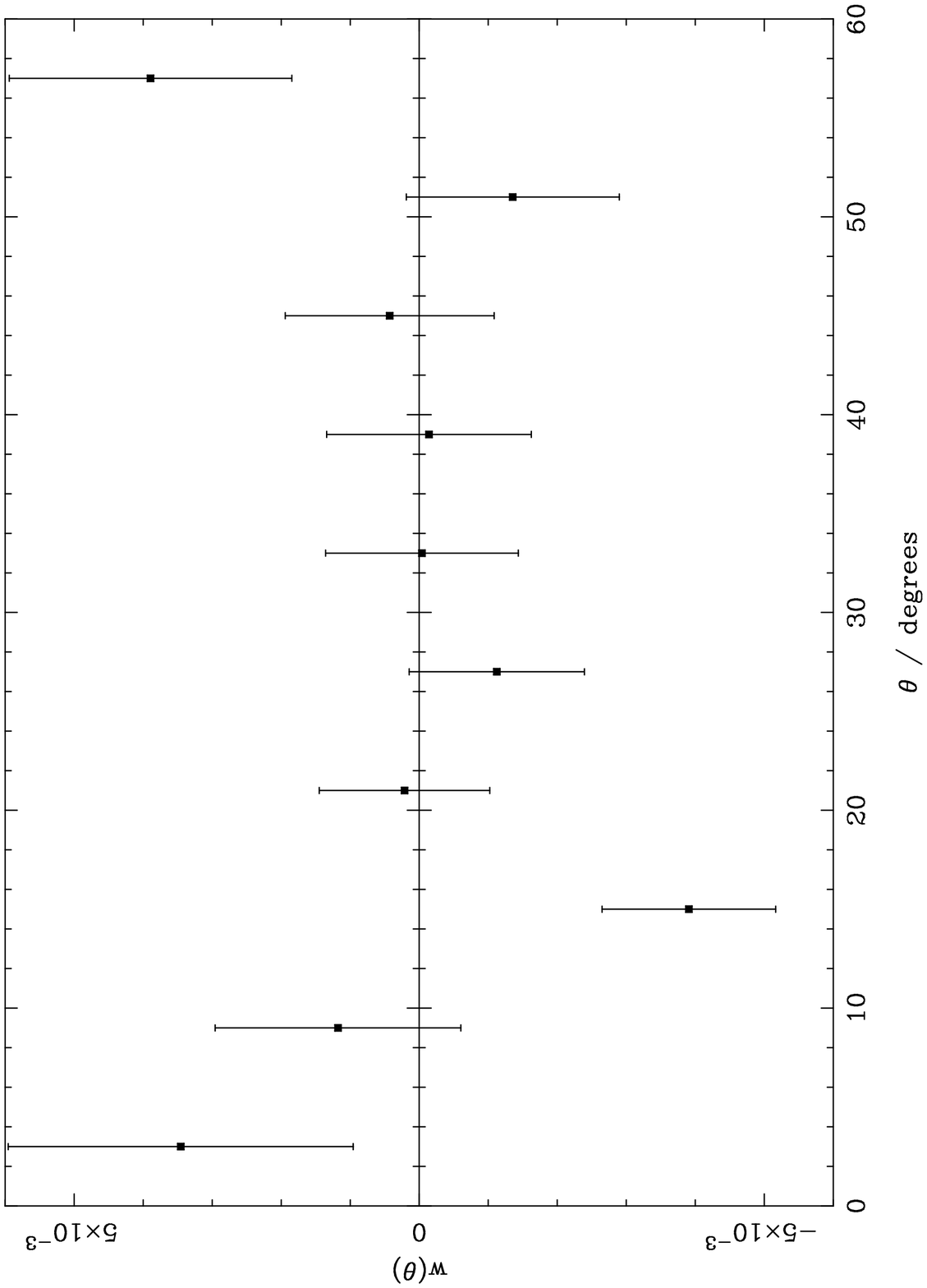}
\caption{\relax}
\end{figure}

% bright:magcor.f; bright:zero7.dat, ccd:satcal.lis (cross-correlation)
\begin{figure}[p]
\epsfbox{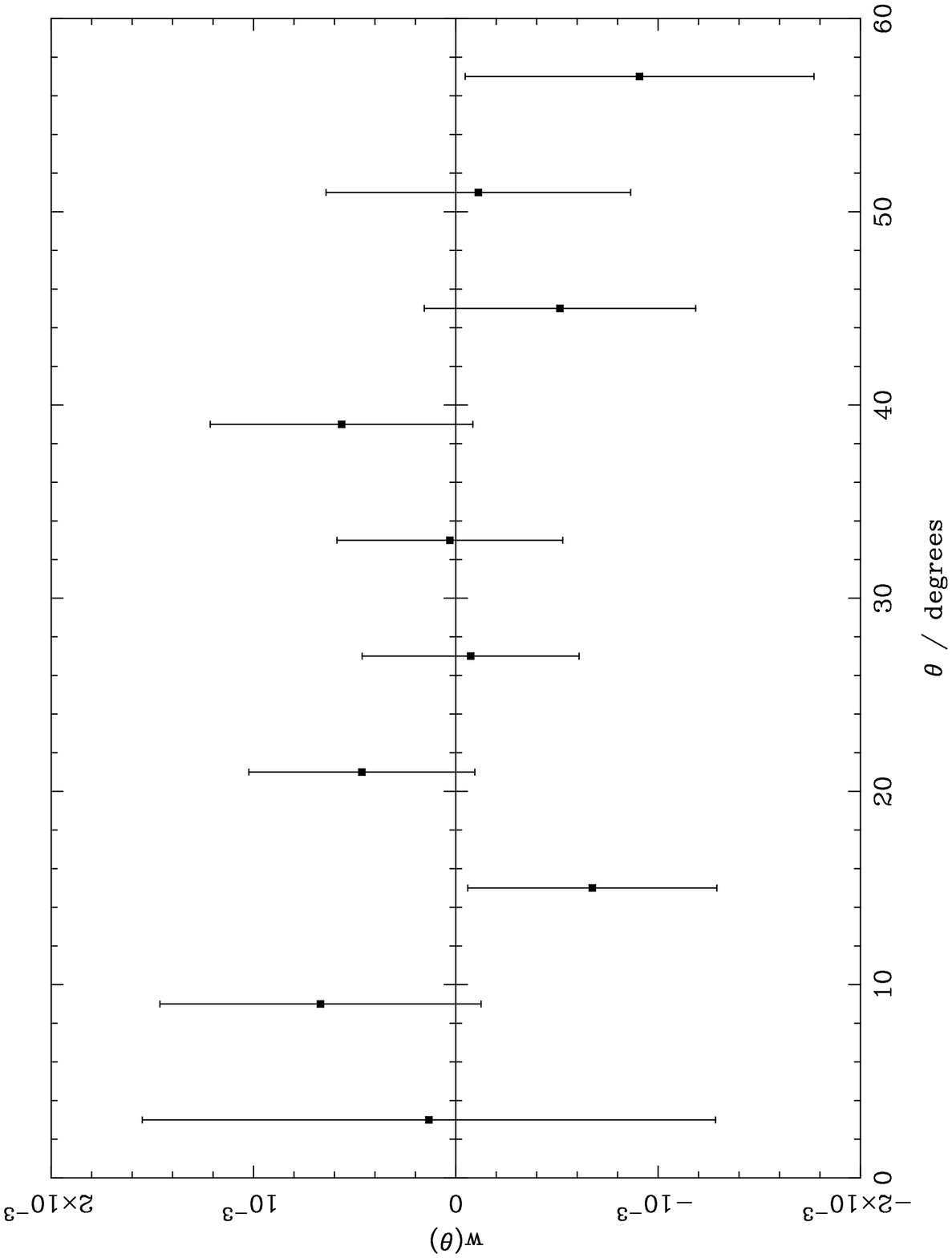}
\caption{\relax}
\end{figure}

% bright:eqa_plot.f; bright:master7.dat
\clearpage
\begin{figure}[p]
\epsfbox{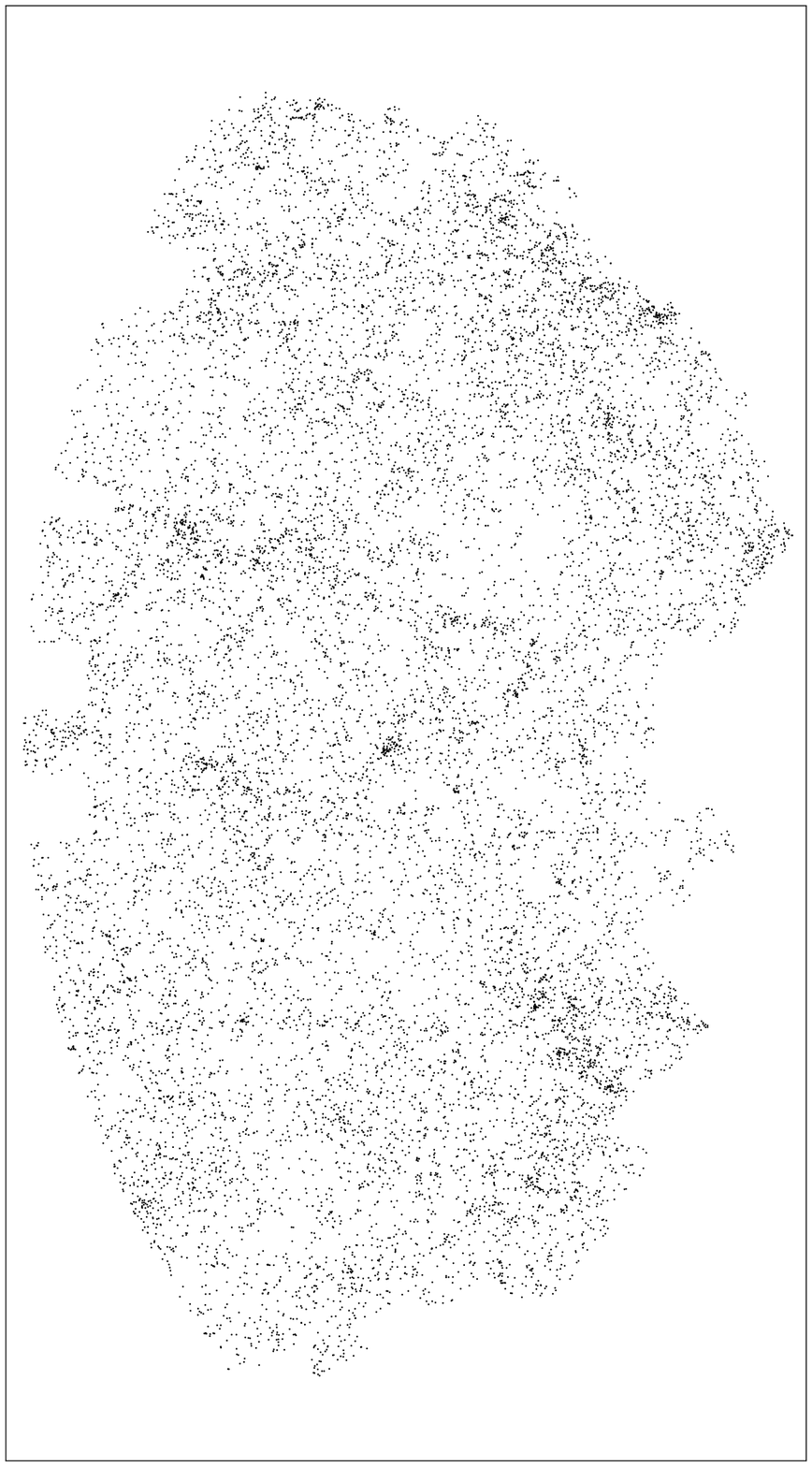}
\caption{\relax}
\end{figure}

% bright:eqa_plot.f; bright:master7.dat
\clearpage
\begin{figure}[p]
\epsfbox{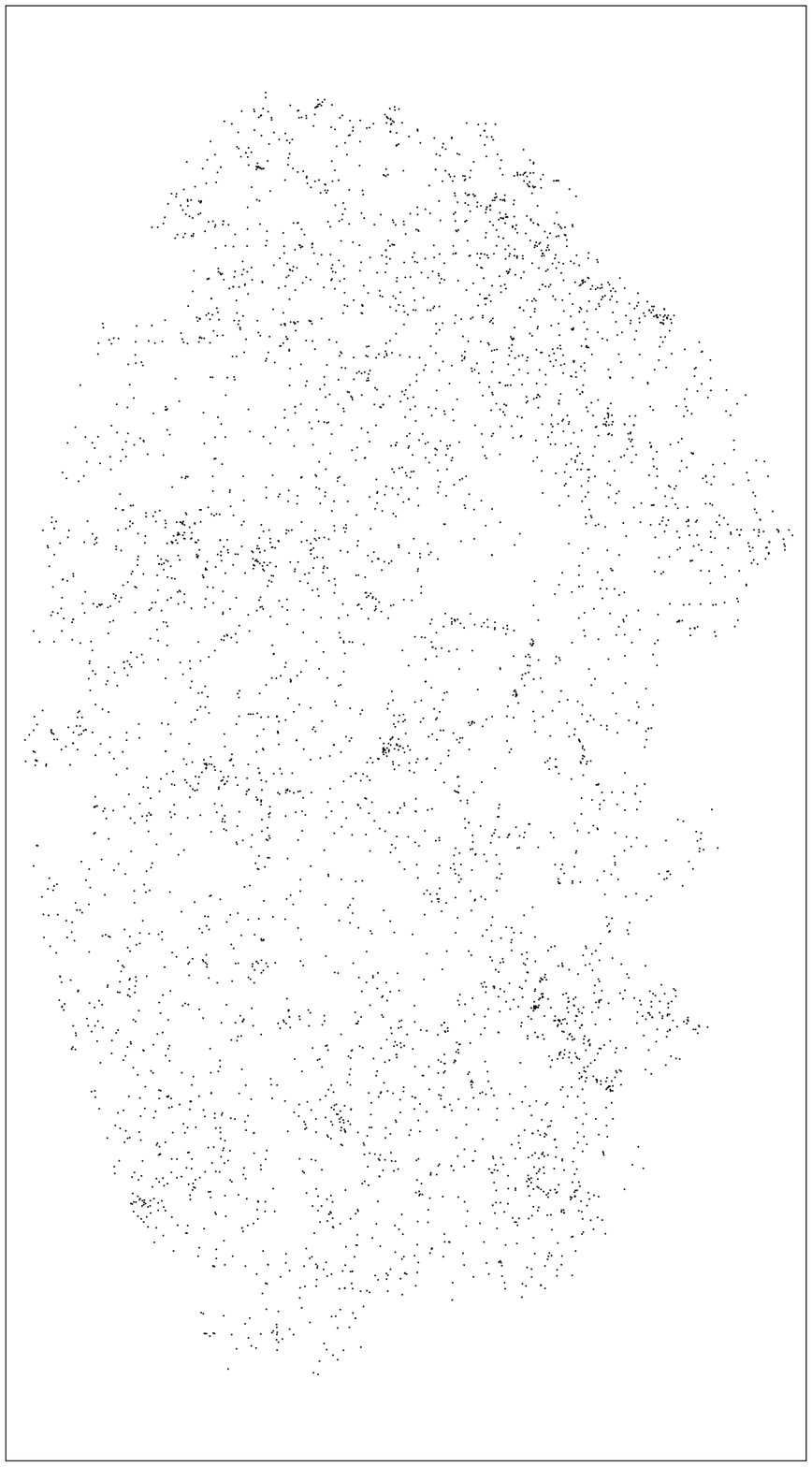}
\caption{\relax}
\end{figure}

% bright:eqa_plot.f; bright:master7.dat
\clearpage
\begin{figure}[p]
\epsfbox{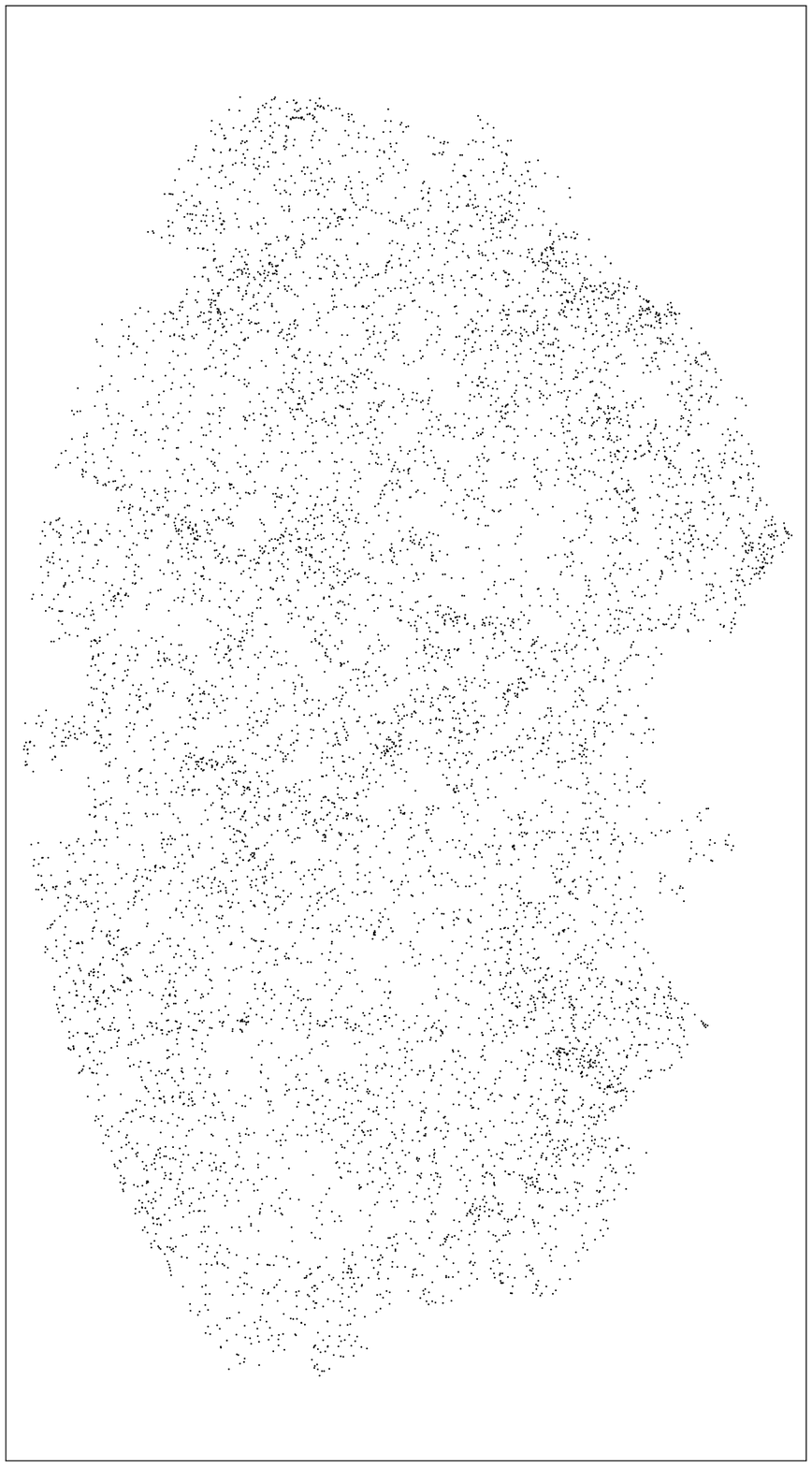}
\caption{\relax}
\end{figure}

% bright:eqa_plot.f; bright:master7.dat
\clearpage
\begin{figure}[p]
\epsfbox{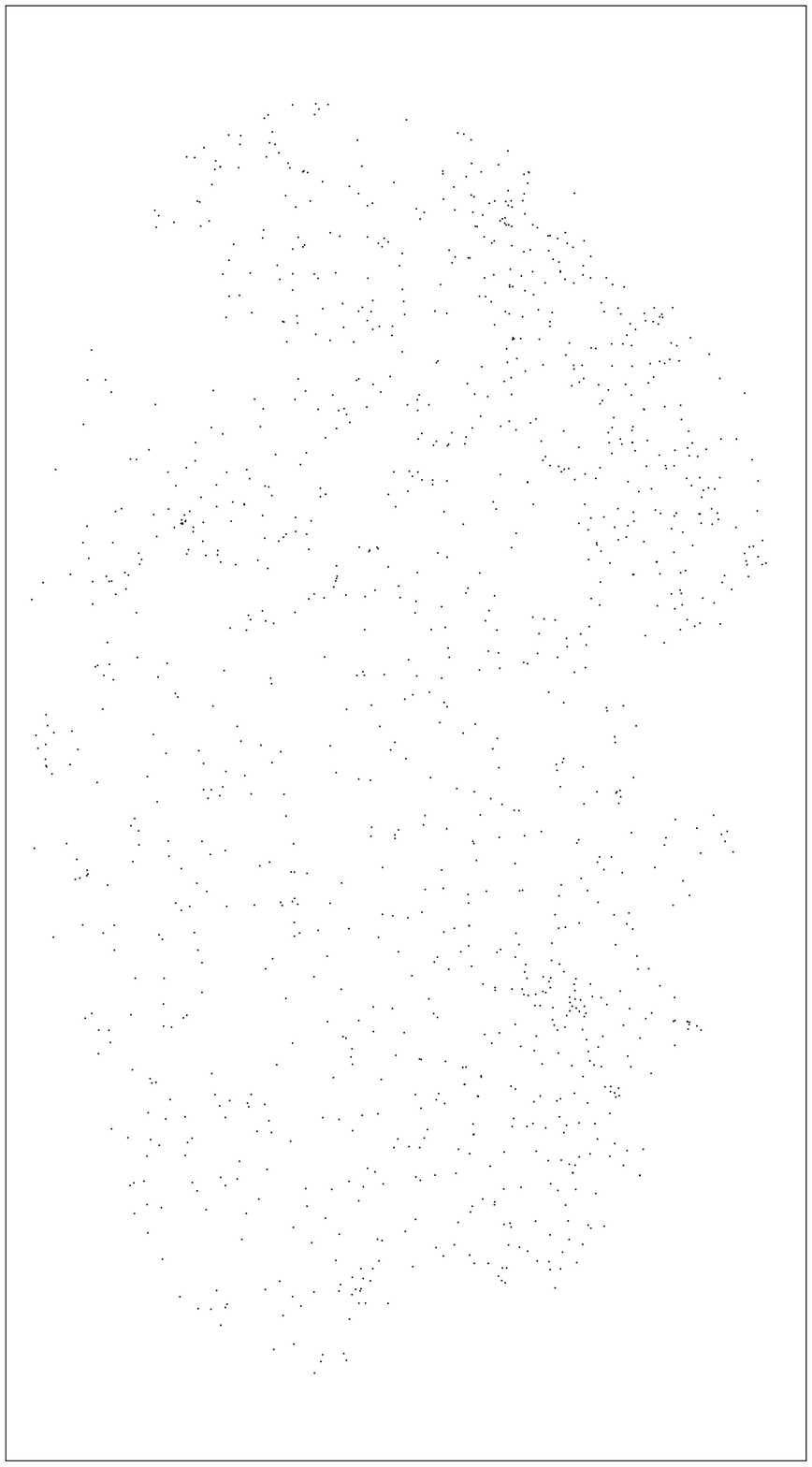}
\caption{\relax}
\end{figure}

% bright:eqa_plot.f; survey:holes.lis
\clearpage
\begin{figure}[p]
\epsfbox{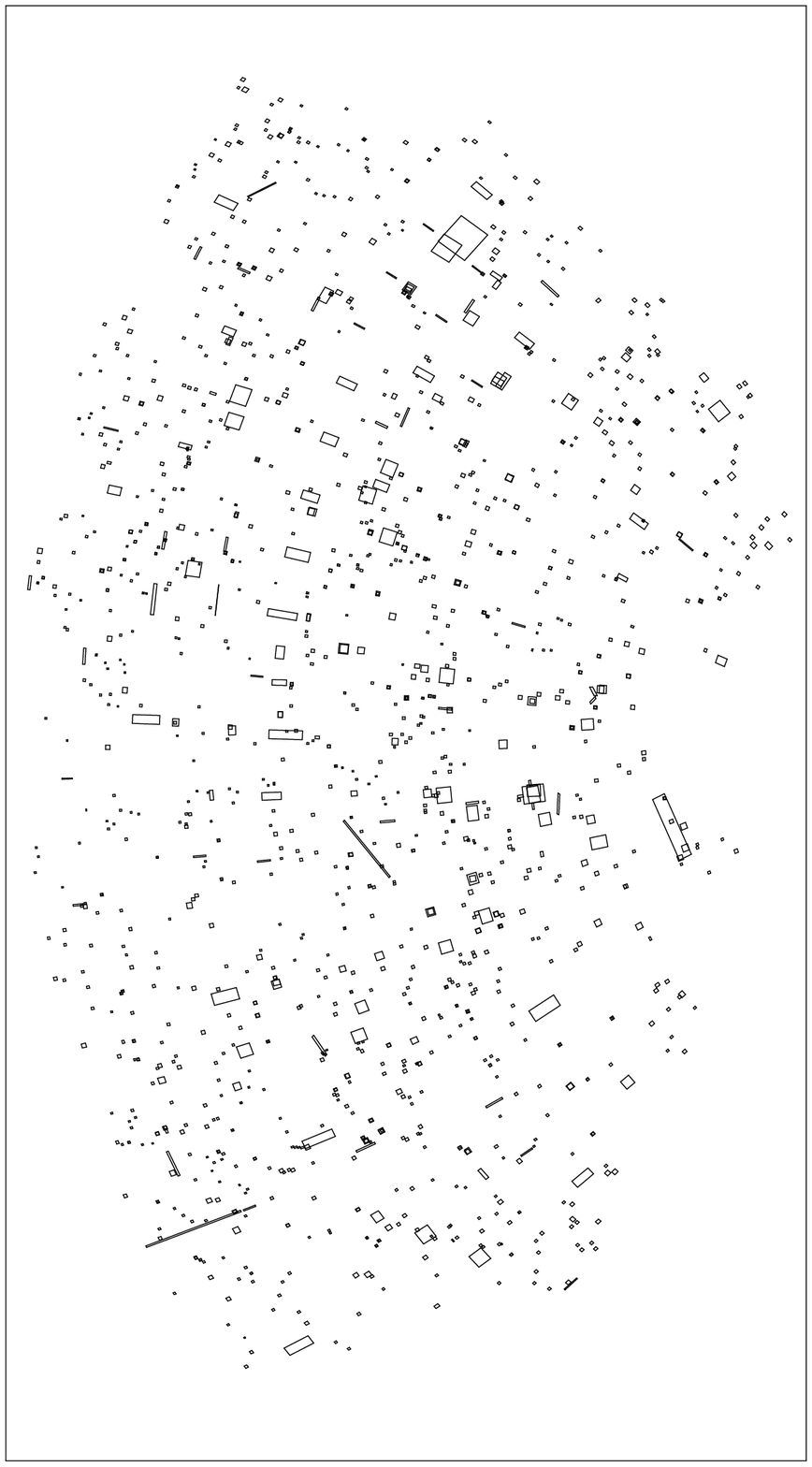}
\caption{\relax}
\end{figure}

% corr:xip2.f; cordat:wp_early_z_n2.dat, wp_all_z_n2.dat, wp_late_z_n2.dat, 
\clearpage
\begin{figure}[p]
\epsfbox{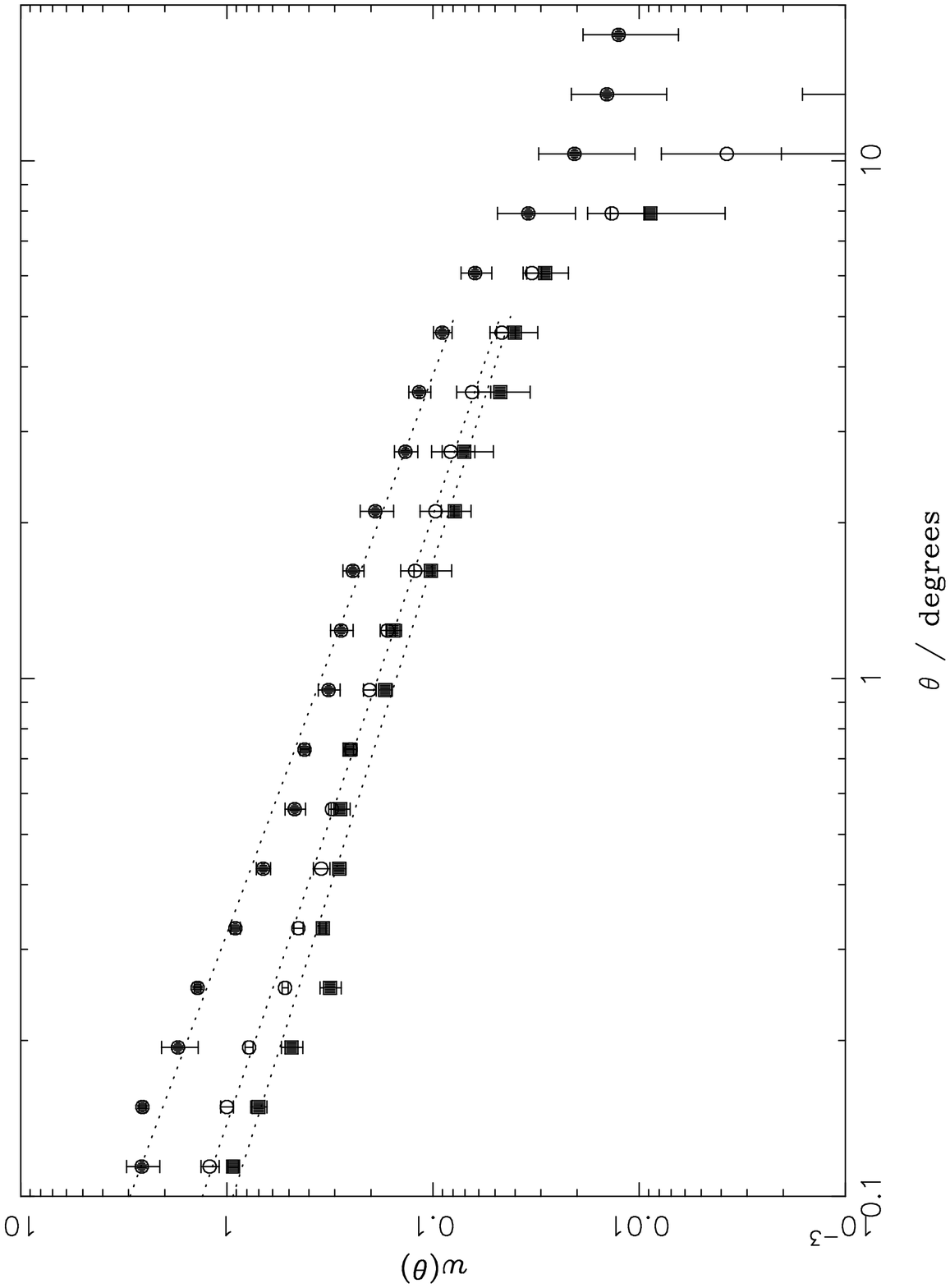}
\caption{\relax}
\end{figure}

\end{document}